\documentclass[
 aps, superscriptaddress,
 amsmath,amssymb,
 reprint,  
]{revtex4-1}
\usepackage[utf8]{inputenc}
\usepackage{graphicx}
\usepackage{dcolumn}
\usepackage{bm}
\usepackage{braket}
\usepackage{verbatim}
\usepackage{lipsum}
\usepackage{amsmath,amssymb}

\usepackage[dvipsnames]{xcolor}

\usepackage{enumitem}
\usepackage{makecell}
\usepackage{relsize}

\usepackage[colorlinks=true, citecolor=Maroon, linkcolor=Bittersweet, urlcolor=RawSienna]{hyperref}

\newcommand{\beq}{\begin{equation}}
\newcommand{\eeq}{\end{equation}}
\newcommand{\beqarray}{\begin{eqnarray}}
\newcommand{\eeqarray}{\end{eqnarray}}

\begin{document}

\title{Coherent and Dynamic Small Polaron Delocalization in CuFeO$_{2}$ }

\author{Jocelyn L. Mendes}
\thanks{These authors contributed equally to this work}
\affiliation{Division of Chemistry and Chemical Engineering, California Institute of Technology, Pasadena, California 91125, USA\looseness=-1}

\author{Srijan Bhattacharyya}
\thanks{These authors contributed equally to this work}
\affiliation{Department of Chemistry, University of Colorado Boulder, Boulder, CO 80309, USA\looseness=-1}

\author{Chengye Huang}
\affiliation{Department of Chemistry, University of California, Berkeley, CA, USA\looseness=-1}

\author{Jonathan M. Michelsen}
\affiliation{Division of Chemistry and Chemical Engineering, California Institute of Technology, Pasadena, California 91125, USA\looseness=-1}

\author{Isabel M. Klein}
\affiliation{Division of Chemistry and Chemical Engineering, California Institute of Technology, Pasadena, California 91125, USA\looseness=-1}

\author{Finn Babbe}
\affiliation{Chemical Sciences Division, Lawrence Berkeley National Laboratory, Berkeley, California 94720, USA\looseness=-1}

\author{Thomas Sayer}
 \affiliation{Department of Chemistry, University of Colorado Boulder, Boulder, CO 80309, USA\looseness=-1}
 \affiliation{Department of Chemistry, Durham University, Durham DH1 3LE, United Kingdom}

\author{Tianchu Li}
 \affiliation{Department of Chemistry, University of Colorado Boulder, Boulder, CO 80309, USA\looseness=-1}

 \author{Jason K. Cooper}
\affiliation{Chemical Sciences Division, Lawrence Berkeley National Laboratory, Berkeley, California 94720, USA\looseness=-1}

\author{Hanzhe Liu}
\email{hanzhe@purdue.edu}
 \affiliation{Department of Chemistry, Purdue University, West Lafayette, Indiana 47907, USA\looseness=-1}

\author{Naomi S. Ginsberg}
\email{nsginsberg@berkeley.edu}
\affiliation{Department of Chemistry, University of California, Berkeley, CA, USA\looseness=-1}
\affiliation{Chemical Sciences Division, Lawrence Berkeley National Laboratory, Berkeley, California 94720, USA\looseness=-1}
\affiliation{Department of Physics, University of California, Berkeley, CA, USA\looseness=-1}
\affiliation{Materials Sciences and Molecular Biophysics and Integrative Bioimaging Divisions, Lawrence Berkeley National Laboratory, Berkeley, California 94720, USA\looseness=-1}
\affiliation{Kavli Energy NanoSciences Institute, Berkeley, California 94720, USA\looseness=-1}
\affiliation{STROBE, NSF Science \& Technology Center, Berkeley, California 94720, USA\looseness=-1}

\author{Andr\'{e}s Montoya-Castillo}
\email{Andres.MontoyaCastillo@colorado.edu}
\affiliation{Department of Chemistry, University of Colorado Boulder, Boulder, CO 80309, USA\looseness=-1}

\author{Scott K. Cushing}
\email{scushing@caltech.edu}
\affiliation{Division of Chemistry and Chemical Engineering, California Institute of Technology, Pasadena, California 91125, USA\looseness=-1}

\date{\today}

\begin{abstract}
Small polarons remain a significant bottleneck in the realization of efficient devices using transition metal oxides. Routes to engineer small polaron coupling to electronic states and lattice modes to control carrier localization remain unclear. Here, we measure the formation of small polarons in CuFeO$_2$ using transient extreme ultraviolet reflection spectroscopy and compare it to theoretical predictions in realistically parameterized Holstein models, demonstrating that polaron localization depends on its coupling to the high-frequency versus low-frequency components of the phonon bath. We measure that small polaron formation occurs on a comparable $\sim$100 fs timescale to other Fe(III) compounds. After formation, a dynamic delocalization of the small polaron occurs through a coherent lattice expansion between Fe-O layers and charge-sharing with surrounding Fe(IV) states. Our simulations of polaron formation dynamics reveal that two major factors dictate polaron formation timescales: phonon density and reorganization energy distributions between acoustic and optical modes, matching experimental findings. Our work provides a detailed, real-time observation of how electronic-structural coupling in a polaron-host material can be leveraged to suppress polaronic effects for various applications.
\end{abstract}

\maketitle


\section*{Introduction}
\vspace{-5pt}
Polarons play a significant role in various material physics, encompassing superconductivity, charge density waves, and electronic transport in both organic and inorganic materials~\cite{Mott1993PolaronSuperconductors, Zhang2021DynamicsOxides, Ghosh2020ExcitonsMaterials, Kudinov2002ModelSuperconductor, Zhang2023BipolaronicSuperconductivity, Nosarzewski2021SuperconductivityModel}. In transition metal oxides with narrow bandwidth, the interplay between strong electron correlations and strong electron-phonon coupling determines properties ranging from band structure to charge transport~\cite{Nosarzewski2021SuperconductivityModel}. For materials with photocatalytic applications, the effect on surface chemical kinetics that benefit from charge localization and lattice reorganization must also be considered~\cite{Yim2016EngineeringSurface, Biswas2019ControllingSpectroscopy}. Iron oxide compounds, such as $\alpha$-Fe$_2$O$_3$ (hematite), are prototypical photocatalysts that possess ideal charge-transfer-based band gaps in the visible spectrum and surface kinetics due to their localized d-orbitals. However, their efficiency falls short of the theoretical maximum because the same properties lead to the formation of photoexcited small polarons, which reduce carrier mobility~\cite{Zhang2019UnderstandingPhotoanodes}. While transition metal oxides that have an empty d-shell after bonding, such as TiO$_2$, tend to form photoexcited large polarons with sufficient carrier mobilities to approach near-theoretical efficiencies, the same bonding also usually leads to large UV bandgaps that limit overall solar absorption efficiency~\cite{Yan2018AnataseTransport}.

The dynamics of small polaron formation are well studied and understood, but the mechanism for delocalizing polarons to improve carrier lifetimes remains unclear~\cite{Biswas2019ControllingSpectroscopy, Carneiro2017Excitation-wavelength-dependent-Fe2O3}. In $\alpha$-Fe$_2$O$_3$, the photoexcited electron distorts the local lattice upon the initial electron-optical phonon scattering, forming small polarons that trap photoexcited carriers with a large carrier effective mass. One proposed approach to suppress small polaron formation is to introduce new atoms at the unit cell level, thereby decreasing the reorganization energy of the small polaron below the thermal energy~\cite{Gregoire2023CombinatorialDiscovery}. Understanding the effect of atomic substitutions on small polaron formation dynamics and their correlation to photoexcited charge transport is necessary for improving photocatalysis and engineering other small-polaron-based applications.

\begin{figure*}[t]
\begin{center} 
\vspace{-3pt}
    \resizebox{.85\textwidth}{!}{\includegraphics[trim={0pt 0pt 0pt 0pt},clip]{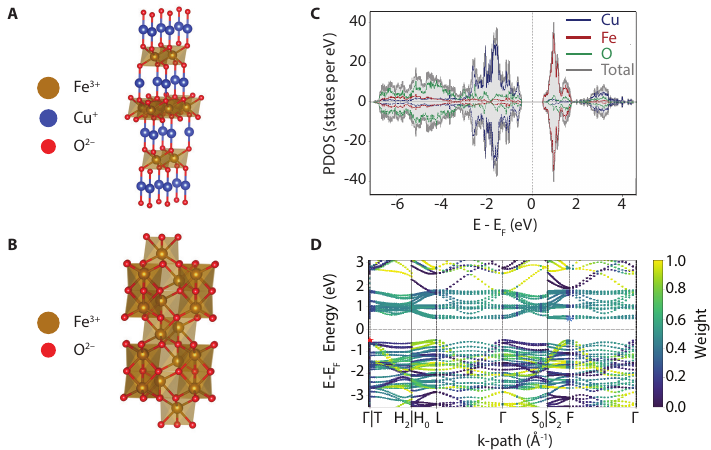}}
    \vspace{-15pt}
\end{center}
\caption{\label{fig1} A comparison of the (A) delafossite crystal structure of CuFeO$_2$ where Cu atoms are inserted between Fe-O layers and (B) hematite ($\alpha$-Fe$_2$O$_3$). (C) The projected density of states shows that the conduction bands are dominated by Fe orbital character while the valence bands are mixed with Cu-O orbitals. (D) The band structure of CuFeO$_2$ -- where red (blue) stars indicate the VBM (CBM). The weights indicate the extent to which each supercell Bloch state contributes to the corresponding primitive cell state.}
\vspace{-5pt}
\end{figure*}

In this work, we measure the delocalization of small polarons in CuFeO$_2$ using transient extreme ultraviolet (XUV) spectroscopy and compare the formation dynamics directly to \textit{ab-initio} calculations, which consider coupling to both the optical and acoustic phonon bath. CuFeO$_2$ has a delafossite structure (Fig.~\ref{fig1}A) with 2D Cu atomic layers substituted between Fe-O networks, and was identified by high-throughput screening techniques as a potential high-performing photocatalyst~\cite{Zhao2019High-ThroughputCalculations, Wolf2023AcceleratedAnalysis, Saadi2006PhotocatalyticCuCrO2}. XUV measurements of CuFeO$_2$ have demonstrated cathodic photocurrent not present in hematite in addition to suggestions of small polaron formation, which are further explored here ~\cite{Husek2018}. The Cu atoms are expected to modulate small polaron formation in two ways: by providing increased electronic screening between photoexcited polarons in the Fe-O layers and by allowing the lattice to expand perpendicular to the Fe-O plane to reduce reorganization potentials. Our transient XUV measurements indicate that electronic screening between layers does not change the small polaron formation rate from the standard $<$100 fs measured for most Fe(III) oxides~\cite{Carneiro2017Excitation-wavelength-dependent-Fe2O3, Husek2017SurfaceFormation, Vura-Weis2013Femtosecondsub3/sub}. Quantum dynamics simulations demonstrate that this polaron formation timescale depends on electronic-structural coupling to different frequency components of the phonon bath. The ability for the lattice to expand perpendicular to the Fe-O plane, is measured to enable a polaron delocalization on a picosecond timescale through a coherent Cu-(Fe-O) c-axis phonon mode~\cite{Klobes2015Anisotropic/math}. A corresponding nearest-neighbor Fe(III) to Fe(IV) charge compensation supports the hypothesis of an initial small-electron-polaron delocalizing beyond the initial Fe(III) site to the nearest neighbors~\cite{deMello1997DynamicalPolarons}. Transient absorption microscopy reveals that the delocalized polaron leads to a longer recombination timescale when compared to $\alpha$-Fe$_2$O$_3$ with similar crystallinity conditions~\cite{Husek2017SurfaceFormation, Fitzmorris2013UltrafastDynamics, Joly2006CarrierReflectivity, Chen2023UltrafastPhotoelectrodes}. Beyond photocatalysis, the results provide insight into a broad range of small polaron based materials and how lattice interactions and electron correlations can be leveraged to tune their properties.

\section*{Results}
\vspace{-5pt}

Figure~\ref{fig1}A shows the crystal structure of CuFeO$_2$, where the Fe-O sublattices are capped by Cu layers along the c-axis. $\alpha$-Fe$_2$O$_3$ is a prototypical example of photoexcited small polaron formation at an Fe(III) site and is shown in Fig.~\ref{fig1}B for reference~\cite{Biswas2019ControllingSpectroscopy, Husek2017SurfaceFormation}. CuFeO$_2$ is a charge transfer insulator (Fig.~\ref{fig1}C) similar to $\alpha$-Fe$_2$O$_3$, in which the valence band maximum (VBM) is dominated by Cu-O hybridization and the conduction band minimum (CBM) is dominated by Fe 3d orbitals. The additional electron density offered by the Cu atoms in the CuFeO$_2$ provides hybridization between the Cu and O atoms, which reduces the material’s band gap compared to $\alpha$-Fe$_2$O$_3$ and increases valence band dispersion (Fig.~\ref{fig1}D). The conduction bands are still relatively flat despite the delocalization offered by the Cu atoms, suggesting that localized electronic states persist in the Fe-O sublattice despite the addition of Cu.

\begin{figure*}[t]
\begin{center} 
\vspace{-3pt}
    \resizebox{.85\textwidth}{!}{\includegraphics[trim={0pt 0pt 0pt 0pt},clip]{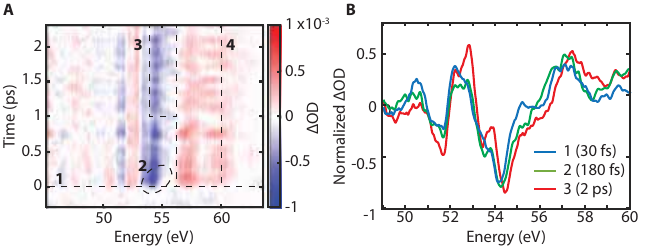}}
    \vspace{-15pt}
\end{center}
\caption{\label{fig2} Transient XUV reflection-absorption spectra of CuFeO$_2$ (A) following photoexcitation with a 400 nm pump. The red and blue colors represent increased and decreased absorption after photoexcitation, respectively. Four main spectral features are labeled in the plot. At pump-probe overlap (t$_0$), a series of positive and negative peaks are observed (label 1). Immediately after photoexcitation, an ultrafast spectral blue shift is observed around 55 eV (label 2), which is attributed to octahedra expansion and small polaron formation. At longer pump-probe time delays, spectral intensity and energy oscillations are observed (label 3), which is attributed to polaron induced coherent acoustic phonons. At higher XUV transition energies, increased XUV absorption is observed and attributed to photoinduced Fe(IV) states (label 4). Experimental lineouts (B) at 30 fs (blue), 180 fs (red), and 2.04 ps (green) which correspond to dynamics in label 1, label 2, and label 3, respectively.}
\vspace{-5pt}
\end{figure*}

Transient XUV spectroscopy is used to measure small polaron formation dynamics in thin film CuFeO$_2$ following a 3.1 eV (400 nm) photoexcitation of the charge transfer transition from the Cu-O orbitals to Fe 3d orbitals. The optical excitation fluence was $\sim$6 mJ/cm$^2$, which results in an initial photoexcited carrier density of $\sim$4.4 $\times$ 10$^{21}$ cm$^{-3}$~\cite{Zurch2017DirectGermanium, Cushing2019DifferentiatingSpectroscopy, deRoulet2024InitialSpectroscopy}. The XUV pulse is generated from few-cycle broadband white light with energy centered around the Fe M$_{2,3}$ edge at $\sim$54 eV. Transient XUV reflection spectroscopy is performed at a 10$^\circ$ grazing incidence angle, resulting in a penetration depth estimated as $\sim$2~nm~\cite{Liu2023MeasuringSpectroscopy}. After photoexcitation, the change in the XUV spectrum is defined as $\Delta OD = -\log_{10} \left( \frac{I_{\text{pump on}}}{I_{\text{pump off}}} \right)
$. The definition is often termed reflection-absorption due to the negative sign. Under this convention, the increase in spectral intensity (red color) in Fig.~\ref{fig2}A is roughly associated with increased absorption, whereas the blue color is roughly associated with decreased absorption. Additional experimental details, including characterization of the CuFeO$_2$ sample, can be found in the Supplementary Materials.

The measured transient XUV spectrum for the Fe M$_{2,3}$ edge is shown in Figure 2. Four distinct spectral features are identified (Fig.~\ref{fig2}A). Immediately following photoexcitation, a series of positive and negative spectral features occur between 51 and 56 eV (Fig.~\ref{fig2}A, label 1). This is followed by a fast spectral blue shift around 55 eV within the first $\sim$100 fs (Fig.~\ref{fig2}A, label 2). There are intensity oscillations in the measured spectra at longer time delays. A clear intensity oscillation is observed for the main negative peak at 55 eV. An increased oscillation in spectral density occurs between 55.5 eV and 56.6 eV, which is out-of-phase compared to the oscillation of the main negative peak between 54.5 eV and 55.5 eV (Fig.~\ref{fig2}A, label 3). In addition to the features between 51 eV and 56 eV, there are two major absorption peaks above 56 eV. These peaks emerge immediately after photoexcitation and exhibit oscillating spectral features that strongly correlate with spectral intensities below 56 eV and are out-of-phase with the oscillations between 54.5 and 55.5 eV. It should be noted that while the features in labels 1 and 2 are common among transient XUV measurements of $\alpha$-Fe$_2$O$_3$ and other prototypical Fe(III) small polaron materials,~\cite{Biswas2019ControllingSpectroscopy, Carneiro2017Excitation-wavelength-dependent-Fe2O3, Kim2024CoherentMechanism, Mendes2025DynamicPolarons} the features in labels 3 and 4 are not typically observed in the transient XUV spectra of photoexcited iron oxides. This suggests that the Cu interstitial layer between the Fe-O octahedra influences longer timescale dynamics.

\begin{figure*}[t]
\begin{center} 
\vspace{-3pt}
    \resizebox{.85\textwidth}{!}{\includegraphics[trim={0pt 0pt 0pt 0pt},clip]{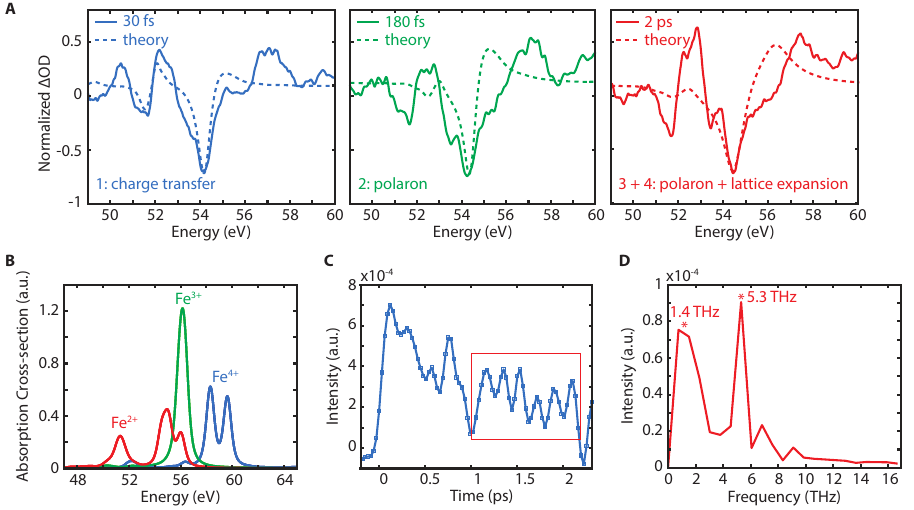}}
    \vspace{-15pt}
\end{center}
\caption{\label{fig3} Experimental lineouts (solid lines) (A) compared to BSE theory (dashed lines) for immediately after photoexcitation (blue), following polaron formation (green), and the change to the polaron spectrum following lattice expansion (red), which correspond the label 1, label 2, and labels 3 and 4 in Fig.~\ref{fig2}A, respectively. Ligand-field theory (B) for differing oxidation states of iron. The blue trace corresponding to Fe(IV) appears at the same energy as Fig.~\ref{fig2}A, label 4. The spectral oscillations (C) in the main negative feature at $\sim$55 eV in Fig.~\ref{fig2}A label 3. The oscillations in the red box were Fourier transformed (d) and demonstrate Cu-(Fe-O) c-axis acoustic modes at 1.4 THz and a mixture of Fe-O modes at 5.3 THz. }
\vspace{-5pt}
\end{figure*}

The observed spectral changes are dominated by changes to the angular momentum coupling X-ray transition Hamiltonian at the Fe M$_{2,3}$ edge, so the measured spectra relate more to oxidation states and local bonding rather than state-filling of photoexcited carriers~\cite{Klein2022iAb/isub3/sub}. To understand the spectral features, we simulate the transient XUV reflection spectra of CuFeO$_2$ using a previously verified adiabatic approximation to an \textit{ab-initio} excited state Bethe-Salpeter equation (BSE) treatment of the XUV spectra (Supplementary Materials). Figure~\ref{fig3}A compares experimental spectral lineouts taken from 30 fs after photoexcitation to 2 ps after photoexcitation (Fig.~\ref{fig3}A) with BSE theory. The lineout immediately following photoexcitation at 30 fs matches a modeled photoexcited charge transfer state. In contrast, the lineout at 180 fs agrees with a modeled small polaron state. Kinetically, the small polaron forms in 90$\pm$20 fs, consistent with previous measurements of Fe(III) oxides~\cite{Husek2017SurfaceFormation, Vura-Weis2013Femtosecondsub3/sub}. The theoretically predicted changes match for both lineouts in terms of the shift of the zero crossing and the appearance of a new multiplet peak at 54.5 eV.

The spectral features in labels 3 and 4 in Fig.~\ref{fig2}A are where the dynamics of CuFeO$_2$ differ from previously measured iron oxide materials. The spectral splitting seen in Fig.~\ref{fig2}A, label 3 matches theory for a c-axis lattice expansion in the DFT unit cell starting around one picosecond after the small polaron has formed (Fig.~\ref{fig3}A, 2 ps, red). When the small polaron is formed, the oxygen octahedra around the new Fe(II) site expand locally due to the quench of the Fe-O bond strength. This secondary, lattice expansion around the polaron site expands the bond length between the oxygen and nearest neighboring Fe(IV) sites through a c-axis expansion, delocalizing the polaron wavefunction from its initial state. The lattice expansion correlates with coherent phonon oscillations measured within this region. Specifically, intensity modulation occurs between 55.4 eV and 56.4 eV, oscillating out of phase compared to the intensity between 54.5 eV and 55.2 eV (Fig.~\ref{fig2}A, labels 3 and 4). The intensity oscillations of the differential spectra between 55.4–56.4 eV is plotted in Fig.~\ref{fig3}C. Initial oscillations are damped due to the optical phonons associated with small polaron formation, but after $\sim$1 ps, a Fourier analysis reveals that dominant oscillations at 1.4 and 5.3 THz are present (Fig.~\ref{fig3}D). These frequencies correspond to mixed acoustic and optical Cu-(Fe-O plane) modes along the \textit{c}-axis and along Fe-O bond lengths, respectively~\cite{Klobes2015Anisotropic/math}. Due to the large anisotropy of the CuFeO$_2$ lattice, phonon modes polarized along the \textit{c}-axis are much softer and anharmonic than those along the \textit{ab}-plane, enabling a low-frequency \textit{c}-axis lattice expansion that coherently couples to localized polaron-induced octahedra expansion.

\begin{figure}[b]
\begin{center} 
\vspace{-3pt}
    \resizebox{.4\textwidth}{!}{\includegraphics[trim={0pt 0pt 0pt 0pt},clip]{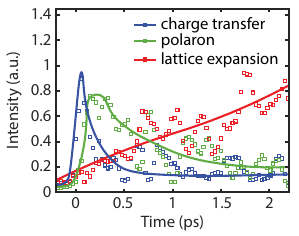}}
    \vspace{-15pt}
\end{center}
\caption{\label{fig4} Kinetics plot of the charge transfer (blue), polaron (green), and lattice expansion (red) states with ``best fit” lines to visualize the kinetics chosen dynamics while disregarding the coherent phonon oscillations.}
\vspace{-5pt}
\end{figure}

\begin{figure*}[t]
\begin{center} 
\vspace{-3pt}
    \resizebox{.85\textwidth}{!}{\includegraphics[trim={0pt 0pt 0pt 0pt},clip]{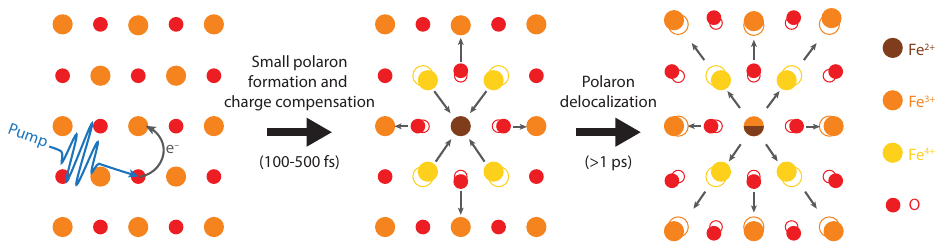}}
    \vspace{-15pt}
\end{center}
\caption{\label{fig5} Schematic representation of the polaron formation mechanism in CuFeO$_2$. The formation of Fe$^{4+}$ states around polaron sites and the expansion of the lattice enables delocalization of the polarons on a ps timescale. }
\vspace{-5pt}
\end{figure*}

The spectral features at energies above 56 eV (Fig.~\ref{fig2}A, label 4) are well represented by the formation of a local Fe(IV) state immediately after photoexcitation which would agree with a delocalization of the Fe(II) small-polaron state through lattice expansion. Modeling local oxidation states in the BSE calculations is challenging, given that carriers are filled up to the photoexcited energy in the adiabatic perturbation limit. To understand oxidation state effects, a ligand-field theory calculation using CTM4XAS is performed and reinforces that the increased XUV absorption above 56 eV is due to Fe(IV) absorption (Fig.~\ref{fig3}B)~\cite{Stavitski2010TheEdges}. This feature also has oscillations that are out-of-phase with the Fe-O and Cu-O bond motions seen in the negative spectral features at 1.4 and 5.3 THz and plotted in (Fig.~\ref{fig3}C and ~\ref{fig3}D). The intrinsic oxidation state in CuFeO$_2$ in the ground state is Fe(III), our measurement suggests that there is a sudden increase in Fe(IV) states after the photoexcited charge-transfer transition. This state then oscillates out-of-phase with the Fe-O and Cu c-axis lattice motions, coupling to the polaronic state and dynamically delocalizing it.

The kinetics for each spectral feature are shown in Fig.~\ref{fig4}. The Fe(IV) state in the nearest-neighbor shell forms immediately after the Fe(III) to Fe(II) charge transfer transition from photoexcitation. However, the polaron delocalization through the oscillation between the ground state bleach and Fe(IV) states does not occur until an order of magnitude longer $\sim$1 ps timescale, compared to the $\sim$100 fs small polaron formation measured here and in other Fe(III) oxide systems~\cite{Husek2017SurfaceFormation, Vura-Weis2013Femtosecondsub3/sub}. The transient XUV data confirm that Cu-hybridization does not alter the small polaron formation energy or kinetics through electronic screening, but rather reduces the on-site reorganization energy, which ultimately delocalizes the small polaron. The small polaron formation and delocalization process are shown schematically in Fig.~\ref{fig5} for CuFeO$_2$. Comparatively, in $\alpha$-Fe$_2$O$_3$, the charge transfer initiated by photoexcitation induces a small electron polaronic state, resulting in lattice distortion and carrier trapping until recombination. In CuFeO$_2$, charge transfer also induces small polaron formation, however, the nearest neighboring atoms compensate for this charge transfer by becoming Fe(IV) in character. The lattice expansion then allows for polaron delocalization on a picosecond timescale.

\begin{figure*}[!t]
\begin{center} 
\vspace{-3pt}
    \resizebox{.7\textwidth}{!}{\includegraphics[trim={0pt 0pt 0pt 0pt},clip]{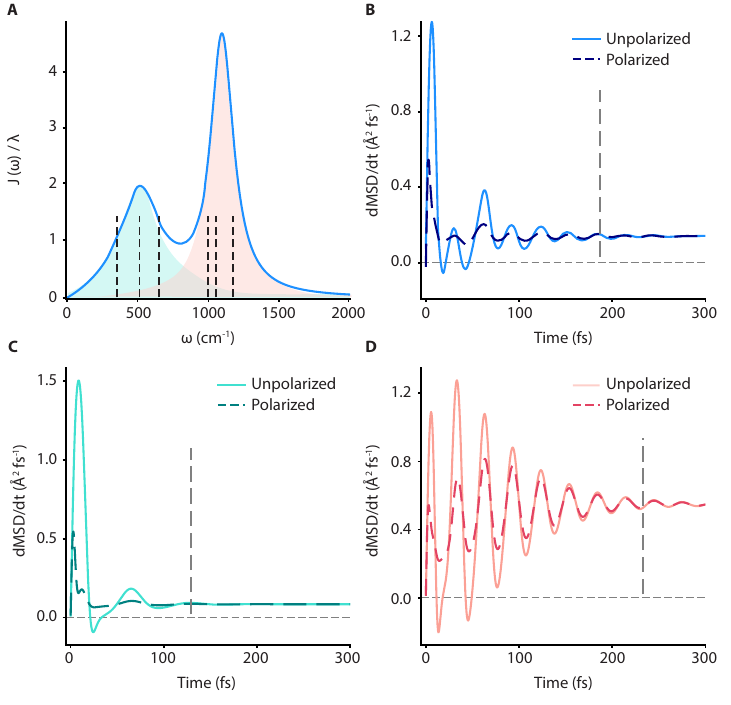}}
    \vspace{-15pt}
\end{center}
\caption{\label{fig6} Tunability of Holstein polaron formation times via the exciton-phonon coupling distributions. (A) Bimodal Lorentzian spectral density based on Raman-detected phonon modes (gray dashed lines). The green and red shaded regions correspond to the low- and high-frequency regions of the full spectral density respectively. (B)-(D) Polaron formation times for exciton-phonon couplings described by: (B) the bimodal distribution in (A), (C) the low-frequency contribution of the bimodal distribution in (A) with the full reorganization energy, and (D) the high-frequency contribution of the bimodal distribution in (A) with the full reorganization energy. (B)-(D) show the derivative of the polaron's mean squared displacement, dMSD/dt, for both \textit{polarized} and \textit{unpolarized} nonequilibrium initial conditions. Gray dashed lines indicate the polaron formation times (i.e., when both nonequilibrium simulations start to agree).}
\vspace{-5pt}
\end{figure*}

To theoretically investigate small polaron formation and its associated kinetics, we employ the dispersive Holstein Hamiltonian~\cite{Holstein1959StudiesMotion}, which encodes the coupling between electronic excitations and local phonons. We parameterize this model using experimental measurements~\cite{Chen2021Temperature-dependentTransition} (i.e., Raman spectra) and electronic structure calculations of the materials of interest~\cite{Ahart2022ElectronTheory, Ferri2019ThermodynamicEnvironments, Ahart2020PolaronicOxides, Naveas2023First-principlesDFT+U+V} (see Supporting Materials). We describe coupling to phonon modes via a continuous spectral density, $J(\omega)$, which we model using two Lorentzians centered on clusters of Raman-active modes around 545 cm$^{-1}$ and 1100 cm$^{-1}$. Figure~\ref{fig6}A shows six peaks taken from the experimental Raman spectrum of CuFeO$_2$ and the resulting bimodal spectral density~\cite{Li2022ADensity}. We further note that, consistent with spectral densities parameterized from molecular dynamics simulations in other materials, one expects the coupling distributions to low-frequency phonons to be broad and high-frequency phonon modes to be narrow~\cite{Zuehlsdorff2019OpticalApproaches, Wiethorn2023BeyondSimulations, Li2023Excitation-Wavelength-DependentDynamics, Wiethorn2024SymmetryBands}. In CuFeO$_2$, the reorganization energy ($\lambda$) that quantifies the total exciton-phonon coupling significantly exceeds the exciton hopping integral ($v$). This allows us to employ the perturbative Non-Interacting Blip Approximation (NIBA)~\cite{Dekker1987Noninteracting-blipBath}, augmented with our recent generalized master equation framework to reach large system sizes free from finite-size effects \cite{Bhattacharyya2025spacelocal}, to predict the quantum dynamics of the polaron, and enabling us to quantify the influence of phonon timescales on the speed of polaron formation.

Because polaron formation timescales are not a direct observable from the quantum dynamics, we extract them indirectly~\cite{Bhattacharyya2024AnomalousBoundaries}. Since the lattice relaxation of polaron formation is a nonequilibrium effect, the time required for the dynamics to become independent of the initial nonequilibrium preparation of the electronic excitation quantifies the polaron formation time. We employ two distinct nonequilibrium preparations of the initial excitation: one where the bare exciton arises from an impulsive Franck-Condon excitation (unpolarized) and is subsequently allowed to move; and one where we allow a static exciton to fully polarize its lattice before it is allowed to move (polarized). We track the deviation between the derivatives of their mean squared displacement (MSD), which encodes polaron mobility, and when these derivatives agree, the polarons have shed their global nonequilibrium preparation, signaling the completion of their local lattice relaxation. We identify the duration of this nonequilibrium relaxation as the polaron formation time~\cite{Bhattacharyya2024AnomalousBoundaries}.

To understand how the form of the spectral density affects the polaron formation time, and therefore how to engineer exciton-phonon couplings to control these, we perform a detailed numerical study (see SI~Sec.~VII). In addition to our earlier insight that increased exciton-nuclear coupling decreases polaron formation times \cite{Bhattacharyya2024AnomalousBoundaries}, we also observe that polaron formation becomes faster when coupled to a \textit{wide and continuous} distribution of phonon frequencies. Further, for a given coupling, these timescales are modulated by the distribution of reorganization energy between the optical and acoustic modes, with increasing weight in the narrow, high-frequency optical modes slowing down polaron formation. Surprisingly, the rate of formation was insensitive to the (mean) vibrational frequency. 

For the spectral density appropriate to CuFeO$_2$, there is a competition of factors, with the larger, high-frequency phonon couplings also having a narrow distribution. Computationally, we unpack the spectral density to understand the relative importance of the low- and high-frequency regions. Keeping the reorganization energy fixed, when considering only the broadly distributed slow (low-frequency) phonons, the timescale of polaron formation is $\sim$130 fs (Fig.~\ref{fig6}C).\footnote{The same distribution without scaling the reorganization energy has a slightly longer timescale of 140~fs, see Fig.~S19.} Contrastingly, the narrowly distributed fast (high-frequency) phonons exhibit a much longer formation timescale of $\sim$230~fs (Fig.~\ref{fig6}D). Together, the full $J(\omega)$ encompassing both regions (Fig.~\ref{fig6}B) gives a value of $\sim$185~fs (which is a weighted combination of the above, see Fig.~S17). We therefore conclude that the addition of narrowly distributed, high-frequency optical phonon modes increases the polaron formation times. Since it is a general feature of condensed phase systems for low-frequency phonon couplings to exhibit spectrally dense, quasi-continuous distributions while couplings to high-frequency phonons are narrowly spectrally distributed, these results are likely applicable beyond transition metal oxides.

\begin{figure*}[t]
\begin{center} 
\vspace{-3pt}
    \resizebox{.85\textwidth}{!}{\includegraphics[trim={0pt 0pt 0pt 0pt},clip]{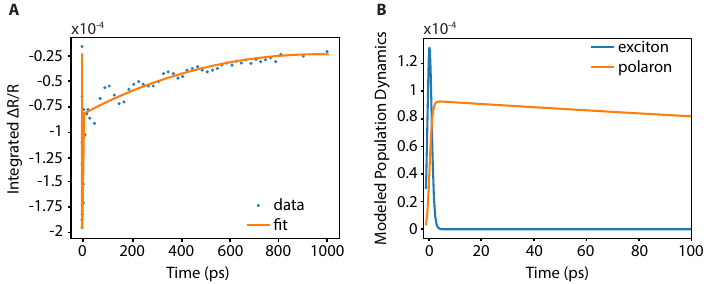}}
    \vspace{-15pt}
\end{center}
\caption{\label{fig7} Optical reflectivity of CuFeO$_2$ (A) and extracted lifetimes of the charge transfer state and polaron state (B). An 8 GHz acoustic phonon oscillation is observed at longer times in the spectra. }
\vspace{-5pt}
\end{figure*}

We further conclude that broadly distributed, low-frequency modes dissipate energy faster and more smoothly than narrowly distributed, high-frequency modes (see SI~Sec.~VII). The smoother response and shorter timescale for low-energy acoustic phonon coupling indicate that the multi-picosecond rise of polaron delocalization in Fig.~\ref{fig4} likely represents the growing occupation of the acoustic phonon bath as the polaron couples to a lattice expansion.

The intersection of experimental and theoretical results also has substantial implications for the description of small polaron physics in transition metal oxides. In particular, the experiment further suggests that, after the polaron is formed, acoustic modes are populated, both in response to the polaron formation and due to the decay of high-energy optical phonons. However, the Holstein model, which offers the standard description of small polaron formation in these materials~\cite{Hannewald2004TheoryCrystals, Ortmann2009TheoryModel, Bhattacharyya2025NonequilibriumDiffusion}, cannot account for such hierarchical energy flow among phonon modes. Instead, the Holstein model offers a field theoretic description of nuclear fluctuations that subsumes energy transfer-facilitating anharmonicities into an effective Gaussian phonon bath. Hence, while Holstein descriptions of small polaron physics can yield useful insights and design principles as we have shown above, precise vibrational engineering further requires the ability to explicitly probe vibrational energy flow in the presence of anharmonic phonons.

Finally, we measure the role of the delocalized polaron state on the carrier lifetime. Interconversion between the polaron and charge transfer state is also examined and confirmed by a variant of optical transient reflectance spectroscopy~\cite{Delor2020ImagingFlow, Weaver2023DetectingNanoscale}. A pump beam of 540~nm center wavelength and $\sim$0.5~ps duration is focused down to a spot with 1/$e^2$ width of 270~nm on the sample to create above-bandgap excitations in the CFO. Subsequently, a wide-field 707~nm probe pulse of similar duration illuminates the sample. The probe reflection is collected by a CMOS camera, and its intensity is integrated over spatial coordinates. The transient reflectance signal is the difference between probe reflectance with and without the pump pulse, scaled by the pump-off reflectance: $\Delta R / R = \frac{I_{\text{pump on}} - I_{\text{pump off}}}{I_{\text{pump off}}}$.

In Fig.~\ref{fig7}, signatures of both the polaron and charge transfer state have negative amplitude but with different strengths. At short time scales ($\sim$ps), a very sharp drop is observed in the transient reflectance signal (Fig.~\ref{fig7}A), corresponding to the fast decay of the charge transfer state as it is converted to the lower amplitude polaron. This longer-lived polaron signal persists for hundreds of ps. The entire kinetic trace is fitted with an instrument response function (0.7~ps) convolved with kinetic equation $A e^{-k t} + B\frac{k}{k_p - k} [e^{-k t} - e^{-k_p t}]$, where A and B are relative amplitudes of the charge transfer and polaron states, respectively. The extracted lifetimes of the charge transfer state ($1/k$) and polaron ($1/k_p$) are 0.52 and 770~ps. The amplitudes of the charge transfer state and polaron have a ratio of $5.6:1$. An acoustic mode at 8~GHz also modulates the signal.

The measured kinetics using transient reflectance agree with the transient XUV findings. Given the temporal resolution of the instrument response function, an initial charge transfer state convolved with a localized polaron persists for picoseconds before transforming into the delocalized polaron state, which then persists on the order of a nanosecond. The finding is intriguing because, although the CuFeO$_2$ is polycrystalline with grain boundaries, it has a lifetime that persists several hundred picoseconds longer than hematite with similar crystallinity conditions~\cite{Fitzmorris2013UltrafastDynamics, Chen2023UltrafastPhotoelectrodes, Joly2006CarrierReflectivity}. The delocalization of the polaron acts to mitigate carrier recombination and extend the lifetime of the localized carriers enabled by population of the acoustic phonon bath. This presents a parameter to tune when engineering polaron-forming semiconductors for photocatalysis, optoelectronics, and other applications.

\section*{Conclusions}
\vspace{-5pt}
The outcomes of this work provide new insight into small polaron-dominated materials and phenomena. We did not find that Cu atom hybridization changed the small polaron formation kinetics, most likely due to strong Fe-Fe electron correlations still dominating interactions. However, changing the lattice degrees of freedom was directly correlated with increasing the delocalization of the small polarons by lowering polaron reorganization energies. Leveraging simulations of lattice descriptions of small polarons in transition metal oxides, we further find that two major factors dictate polaron formation timescales: spectral phonon density and reorganization energy distributions between acoustic and optical modes. Specifically, we find that phonon bath engineering favoring strong coupling to broad acoustic phonon distributions in favor of narrowly distributed optical modes decreases polaron formation timescales. We further find that polaron delocalization is a dynamical process that occurs after the formation of small polarons, contrasting with the concept of immediate delocalized polaron formation~\cite{Kim2024CoherentMechanism}. Furthermore, the delocalization of the polaron appears to be related to longer photoexcited lifetimes in the material when compared to prototypical hematite. This work provides insights into more mobile polarons in desirable transition metal oxides, as well insights into coherent control of the lattice response to small polaron formation.

\section*{Acknowledgments}
\vspace{-5pt}

This material is based on work performed by the Liquid Sunlight Alliance, which is supported by the U.S. Department of Energy, Office of Science, Office of Basic Energy Sciences, Fuels from Sunlight Hub under Award Number DE-SC0021266. Ground state optical reflectivity data were collected at the Molecular Materials Research Center in the Beckman Institute of the California Institute of Technology. The transient optical studies were partially supported by the Center for Computational Study of Excited State Phenomena in Energy Materials (grant no. C2SEPEM) under the US Department of Energy, Office of Science, Basic Energy Sciences, Materials Sciences and Engineering Division (contract no. DE-AC02-05CH11231), as part of the Computational Materials Sciences Program. The computations presented here were partially conducted in the Resnick High Performance Computing Center, a facility supported by the Resnick Sustainability Institute at the California Institute of Technology. This work also partially utilized the Alpine high-performance computing resource at the University of Colorado Boulder. Alpine is jointly funded by the University of Colorado Boulder, the University of Colorado Anschutz, Colorado State University, and the National Science Foundation (award 2201538)

I.M.K and J.L.M.~acknowledge support by the National Science Foundation Graduate Research Fellowship Program under Grant No. 1745301. S.B., T.L., and A.M.C.~were supported by the National Science Foundation Early Career Award in the directorate for Mathematical and Physical Sciences under Award No.~2443961. A.M.C. and N.S.G.~acknowledge the support from a David and Lucile Packard Fellowship for Science and Engineering. N.S.G.~also acknowledges the support of the STROBE Center for Realtime Imaging, a National Science Foundation Science and Technology Center (grant DMR 1548924).  S.B.~thanks Ming-Chun Jiang for insightful discussions on the electron structure calculations. T.S.~is the recipient of an Early Career Fellowship from the Leverhulme Trust. Any opinions, findings, and conclusions or recommendations expressed in this material are those of the author(s) and do not necessarily reflect the views of the National Science Foundation.


\bibliography{ref_main.bib}

\end{document}



\setcounter{section}{0}
\setcounter{equation}{0}
\setcounter{figure}{0}
\setcounter{table}{0}
\setcounter{page}{1}

\renewcommand{\theequation}{S\arabic{equation}}
\renewcommand{\thefigure}{S\arabic{figure}}
\renewcommand{\thepage}{S\arabic{page}}
\renewcommand{\bibnumfmt}[1]{$^{\mathrm{S#1}}$}
\renewcommand{\citenumfont}[1]{S#1}

\title{Supplemental information for ``Coherent and Dynamic Small Polaron Delocalization in CuFeO$_{2}$ ''}

\author{Jocelyn L. Mendes}
\thanks{These authors have contributed equally to this work}
\affiliation{Division of Chemistry and Chemical Engineering, California Institute of Technology, Pasadena, California 91125, USA\looseness=-1}

\author{Srijan Bhattacharyya}
\thanks{These authors have contributed equally to this work}
\affiliation{Department of Chemistry, University of Colorado Boulder, Boulder, CO 80309, USA\looseness=-1}

\author{Chengye Huang}
\affiliation{Department of Chemistry, University of California, Berkeley, CA, USA\looseness=-1}

\author{Jonathan M. Michelsen}
\affiliation{Division of Chemistry and Chemical Engineering, California Institute of Technology, Pasadena, California 91125, USA\looseness=-1}

\author{Finn Babbe}
\affiliation{Chemical Sciences Division, Lawrence Berkeley National Laboratory, Berkeley, California 94720, USA\looseness=-1}

\author{Isabel M. Klein}
\affiliation{Division of Chemistry and Chemical Engineering, California Institute of Technology, Pasadena, California 91125, USA\looseness=-1}

\author{Thomas Sayer}
 \affiliation{Department of Chemistry, University of Colorado Boulder, Boulder, CO 80309, USA\looseness=-1}
 \affiliation{Department of Chemistry, Durham University, Durham DH1 3LE, United Kingdom}

\author{Tianchu Li}
 \affiliation{Department of Chemistry, University of Colorado Boulder, Boulder, CO 80309, USA\looseness=-1}

\author{Jason K. Cooper}
\affiliation{Chemical Sciences Division, Lawrence Berkeley National Laboratory, Berkeley, California 94720, USA\looseness=-1}

\author{Hanzhe Liu}
\homepage{hanzhe@purdue.edu}
 \affiliation{Department of Chemistry, Purdue University, West Lafayette, Indiana 47907, USA\looseness=-1}

\author{Naomi S. Ginsberg}
\homepage{nsginsberg@berkeley.edu}
\affiliation{Department of Chemistry, University of California, Berkeley, CA, USA\looseness=-1}
\affiliation{Chemical Sciences Division, Lawrence Berkeley National Laboratory, Berkeley, California 94720, USA\looseness=-1}
\affiliation{Department of Physics, University of California, Berkeley, CA, USA\looseness=-1}
\affiliation{Materials Sciences and Molecular Biophysics and Integrative Bioimaging Divisions, Lawrence Berkeley National Laboratory, Berkeley, California 94720, USA\looseness=-1}
\affiliation{Kavli Energy NanoSciences Institute, Berkeley, California 94720, USA\looseness=-1}
\affiliation{STROBE, NSF Science \& Technology Center, Berkeley, California 94720, USA\looseness=-1}

\author{Andr\'{e}s Montoya-Castillo}
\homepage{Andres.MontoyaCastillo@colorado.edu}
\affiliation{Department of Chemistry, University of Colorado Boulder, Boulder, CO 80309, USA\looseness=-1}

\author{Scott K. Cushing}
\homepage{scushing@caltech.edu}
\affiliation{Division of Chemistry and Chemical Engineering, California Institute of Technology, Pasadena, California 91125, USA\looseness=-1}

\maketitle

\tableofcontents
\newpage

\newpage

\onecolumngrid

\section{Material Synthesis and Characterization}

\subsection{Preparation of the CuFeO\texorpdfstring{$_{2}$}{2} Thin Films}
Thin-film samples of rhombohedral CuFeO2 were prepared using magnetron co-sputtering in a Lesker LAB Line sputtering system. The samples were deposited on quartz substrates (25 $\times$ 75 mm$^2$, CGQ-0640-01, Chemglass) that were cleaned beforehand. The sputtering process used 2$^{\prime\prime}$ Cu and Fe targets with RF powers of 34~W and 136~W, respectively. A reactive atmosphere of 10\% oxygen in argon at a pressure of 10~mTorr was used. During the 30-minute deposition step, samples are heated to 200$^\circ$~C to aid the crystallization process. Following deposition, the samples underwent a 6-hour post-annealing process at 600$^\circ$~C in a quartz glass tube furnace under a flowing argon atmosphere (100~sccm).

The metal composition ratio is determined using inductively coupled plasma mass spectroscopy (Agilent 7900 ICP-MS). For this samples are digested in nitric acid and diluted before injection. The overall Cu/Fe ratio is determined to be close to stoichiometric, with 1.03. The sample thickness is determined to be 200.4$\pm$0.1~nm via spectroscopic ellipsometry.

\subsection{Ground State Optical Characterization}
\begin{figure}[!h]
\vspace{-6pt}
\begin{center} 
    \resizebox{.95\textwidth}{!}{\includegraphics{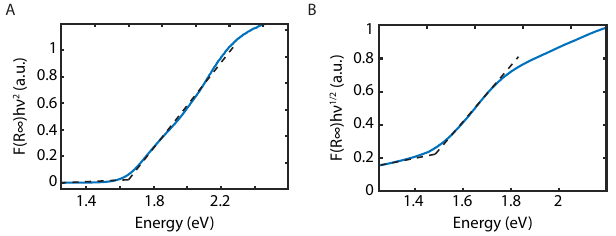}}
\end{center}
\vspace{-24pt}
\caption{\label{fig:s1} Tauc plot of the ground state reflectivity of the polycrystalline CuFeO$_{2}$ thin film for direct allowed (A) and indirect allowed (B) electronic transitions.}
\vspace{-4pt}
\end{figure}

The optical band gap of thin film CuFeO$_{2}$ was measured using UV-Visible reflectance spectroscopy (Cary 5000, Agilent Technologies). Figure.~\ref{fig:s1} presents an $(\alpha h \nu)^{1/r}$ vs energy plot where $r = \dfrac{1}{2}$ for direct allowed (Fig.~\ref{fig:s1}A) and $r = 2$ for indirect allowed (Fig.~\ref{fig:s1}B) energy transitions. For a Tauc plot analysis, the reflectance spectra were transformed using the Kubelka-Munk function using Eq~\ref{SI:eq1}:
\begin{equation}\label{SI:eq1}
 F(R_{\infty}) = \frac{(1-R_{\infty})^2}{2 R_{\infty}},
\end{equation}
where  $R_{\infty} = R_{\rm sample}/ R_{\rm standard}$~\cite{Makua2018HowSpectra}. The direct band gap is estimated to be ~1.63 eV and the indirect band gap is estimated to be $\sim1.48$ eV.

\subsection{Structural Characterization}

\begin{figure}[!h]
\vspace{-6pt}
\begin{center} 
    \resizebox{.5\textwidth}{!}{\includegraphics{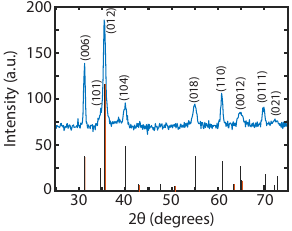}}
\end{center}
\vspace{-24pt}
\caption{\label{fig:s2} Experimental (blue) and calculated (black) XRD spectrum of thin film CuFeO$_{2}$.}
\vspace{-4pt}
\end{figure}

The crystal structure of the CuFeO\textsubscript{2} thin film was characterized via X-ray diffraction (XRD). Fig.~\ref{fig:s2} details the experimental XRD trace (blue) and calculated XRD features (black) from the ICDD database (ICSD \#01-075-2146). When comparing the experimentally measured diffraction pattern against a simulated pattern we find that the peak positions agree and the CuFeO\textsubscript{2} film adopts the expected rhombohedral crystal structure.

\begin{figure}[!h]
\vspace{-6pt}
\begin{center} 
    \resizebox{.5\textwidth}{!}{\includegraphics{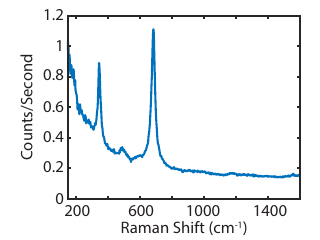}}
\end{center}
\vspace{-24pt}
\caption{\label{fig:s3} Raman shift of CuFeO\textsubscript{2}.}
\vspace{-4pt}
\end{figure}

Further structural characterization was achieved by collecting Raman spectra of the CuFeO\textsubscript{2} thin films. The E\textsubscript{g} and A\textsubscript{1g} modes previously reported for CuFeO\textsubscript{2} are observed at 348 and 688~cm\textsuperscript{-1}, respectively (Fig.~\ref{fig:s3})~\cite{Pavunny2010RamanCuFeO2}.

\section{Experimental setup: Transient Extreme Ultraviolet Reflectivity Spectrometer}

The transient extreme ultraviolet (XUV) reflectivity spectrometer has been described previously~\cite{Liu2023MeasuringSpectroscopy, Kim2024CoherentMechanism}. Briefly, the reported thin film CuFeO\textsubscript{2} measurements were obtained using photoexcitation from a $\sim$35~fs, 400~nm frequency-doubled output of a BBO crystal with \textit{p}-polarization, pumped by an 800~nm, 1~kHz regeneratively amplified Ti:Sapphire laser (Coherent Legend Elite Duo). The optical excitation fluence was approximately 6~mJ/cm\textsuperscript{2}, resulting in an initial photoexcited carrier density of $\sim$4.4~$\times$~10\textsuperscript{21}~cm\textsuperscript{--3}~\cite{deRoulet2024InitialSpectroscopy, Zurch2017DirectGermanium, Cushing2019DifferentiatingSpectroscopy}.  Transient reflection was measured by varying the delay times between the pump and probe pulses using an optomechanical delay stage on the pump line. The XUV probe pulse was generated by high-harmonic generation in argon gas using an \textit{s}-polarized few-cycle white light beam ($<$6~fs, 550--950~nm), characterized previously~\cite{Liu2023MeasuringSpectroscopy}. A 200~nm thick Al filter (Luxel) was used to remove the fundamental white light beam. The resulting XUV continuum probed the Fe M\textsubscript{2,3} absorption edge around 54~eV. A 10$^\circ$ grazing incidence reflection geometry (80$^\circ$ from normal incidence) was employed. Edge-pixel referencing based on signal-free spectral regions was used to reduce noise due to intensity fluctuations~\cite{Geneaux2021SourceReferencing}.

\section{Experimental Setup: Transient optical reflectance spectroscopy}
A 540~nm pump pulse was generated by a Light Conversion non-collinear optical parametric amplifier (model PN13F1) using the third harmonic of the fundamental 1080~nm seed pulse from a Light Conversion Pharos laser (model 10-200-PP). A 707~nm probe pulse was generated by a Light Conversion non-collinear optical parametric amplifier (model PN15F2) using the second harmonic of the fundamental seed pulse. The laser repetition rate was 200~kHz; the pump was modulated at 660~Hz, and the pump--probe delay times were controlled by a mechanical stage.

The pump and probe beams were spatially filtered through 25~$\mu$m and 50~$\mu$m pinholes, respectively. The pump beam was telescoped to a diameter of $\sim$12~mm, while the probe beam was telescoped to 1~mm and focused into the back focal plane of the objective of a home-built microscope using an $f = 300$~mm lens. The two beams were combined using a dichroic mirror (DMLP505, Thorlabs), and a 50/50 beamsplitter reflected both beams into a high numerical aperture (1.4~NA) oil-immersion objective (Leica HC PL APO 63$\times$/1.40~NA) and onto the sample, resulting in overlapped confocal pump and wide-field probe illumination. Probe light reflected from the sample--substrate interface, as well as scattered from the sample, was collected through the same objective. The probe light was isolated with a long-pass filter (FEL605, Thorlabs) and focused onto a charged metal oxide semiconductor (CMOS) detector with 5.86~$\mu$m square pixels, triggered at 660~Hz (PixeLINK PL-D752, equipped with a Sony IMX174 global shutter sensor), using an $f = 500$~mm lens placed one tube length (200~mm) away from the back focal plane of the objective. The total magnification is calculated as $63 \times 500/200 = 157.5$, giving a spatial resolution of 37.2~nm/pixel. Data were cropped to maintain a field of view of 4~$\mu$m~$\times$~4~$\mu$m. The transient reflectance signal was generated by taking the difference in integrated signal between pump-on and pump-off frames, normalized to the integrated pump-off intensities, yielding $\Delta R/R$. Setup automation and data acquisition were implemented in LabVIEW 2014 (64-bit). Data analysis and plotting were performed using a combination of ImageJ (Fiji) and Python. The pump fluence (0.42~mJ/cm\textsuperscript{2}) was chosen to remain within a regime where the spatially integrated signal decay was independent of pump power, corresponding to a carrier density of $\sim$10\textsuperscript{19}~cm\textsuperscript{--3}.

In fitting the population dynamics, the entire trace was modeled by convolving the decay dynamics with the instrument response function. The fitting parameters are summarized below in Table.~\ref{table0}.

\begin{table}[!h]
    \centering
    \vspace{-1.5pt}
    \hspace{-3pt}
    \begin{tabular}{cc} 
    \hline
   Parameter & Value \\ 
     \hline
    $A$ & $-5.2 \times 10^{-4}$ \\ \hline
    $B$ & $-9.3 \times 10^{-5}$\\ \hline
    $k$ & 1.9 $\rm ps^{-1}$  \\ \hline
    $k_{p}$ & $1.3 \times 10^{-3}$ \rm ps$^{-1}$\\ \hline
    \end{tabular}
    
    \caption{\label{table0} Fitting parameters of the CuFeO$_{2}$ population dynamics}
    \vspace{-4pt}
\end{table}

\section{Electronic Structure Calculation}

\begin{figure}[!b]
\vspace{-6pt}
\begin{center} 
    \resizebox{.8\textwidth}{!}{\includegraphics{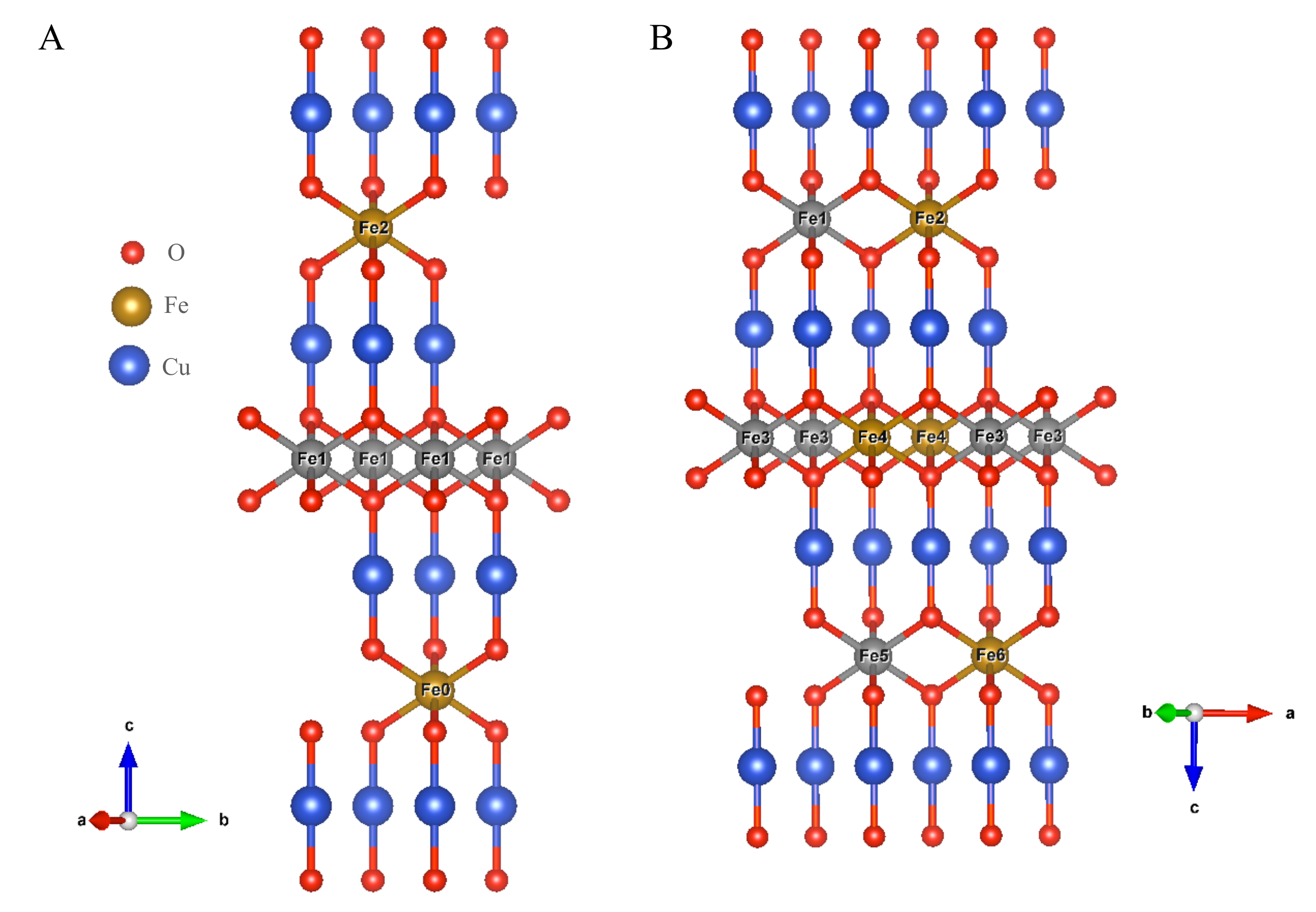}}
\end{center}
\vspace{-24pt}
\caption{\label{fig:crystal} Crystal structure of CuFeO$_2$ in the primitive cell (A) and supercell (B).  Gold and Grey spheres represent Fe atoms but with different spin polarization. Cu atoms and O atoms are represented in blue and red respectively.}
\vspace{-4pt}
\end{figure}

To investigate the electronic structure of CuFeO$_{2}$, we adopted a primitive cell crystal structure from the Materials Project database (Materials Project ID: mp-510281). We computed the total density of states (DOS), projected density of states (PDOS), and band structure calculations using density functional theory (DFT) with the Perdew–Burke–Ernzerhof (PBE) generalized gradient approximation (GGA) for the exchange-correlation functional~\cite{Perdew1996GeneralizedSimple}, along with projector augmented-wave (PAW) pseudopotentials~\cite{Blochl1994ProjectorMethod}. The primitive cell is hexagonal, with lattice parameters: $a = b = 3.0709$ \AA, $c = 17.2513$ \AA, and angles $\alpha = \beta = 90^\circ$, $\gamma = 120^\circ$. This unit cell contains 12 atoms in total, comprising three distinct Fe atoms, three Cu atoms, and six O atoms.

\begin{figure}[!t]
\vspace{-6pt}
\begin{center} 
    \resizebox{.95\textwidth}{!}{\includegraphics{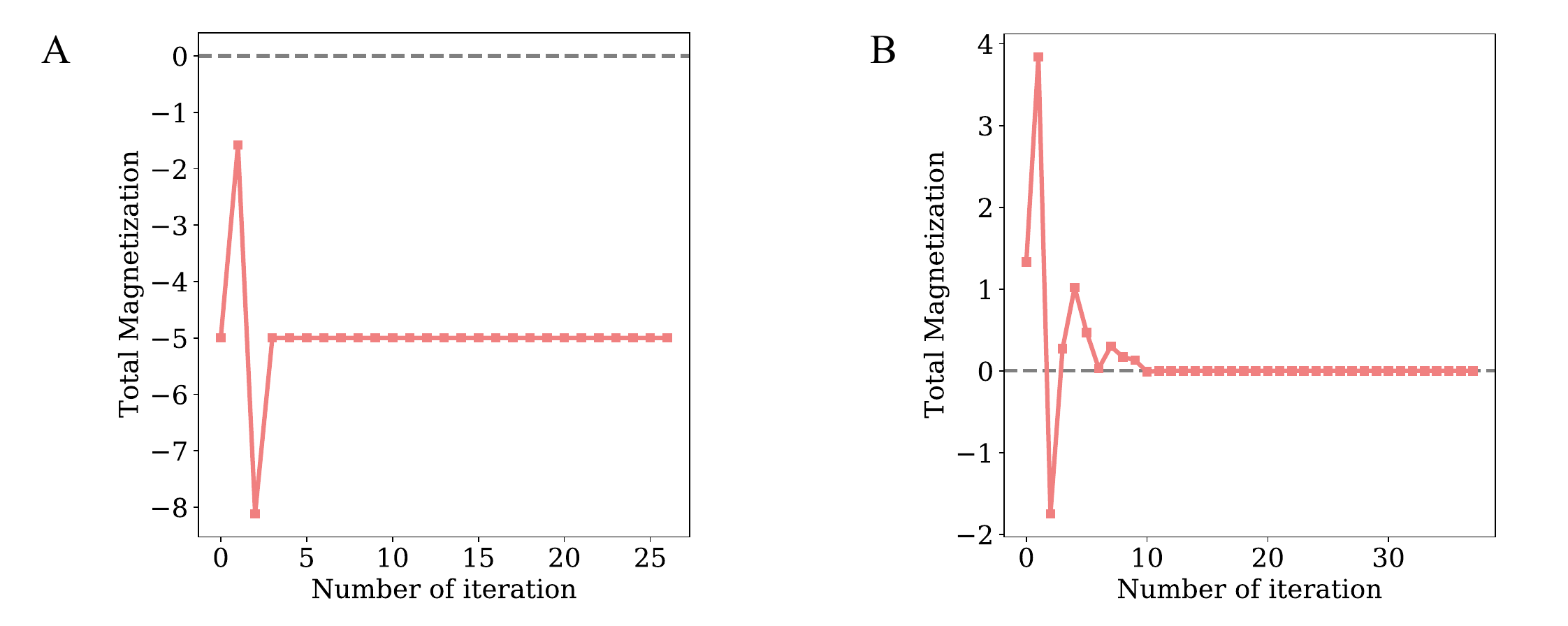}}
\end{center}
\vspace{-24pt}
\caption{\label{fig:magnetization} Total magnetization of CuFeO$_2$ as a function of number of SCF iterations in the primitive cell (A) and $2 \times 1 \times 1$ supercell (B).}
\vspace{-4pt}
\end{figure}

Before discussing the electronic properties of CuFeO$_2$, one must first establish the correct magnetic ordering. In the primitive cell, three crystallographically distinct Fe atoms are present. Given that CuFeO$_2$ is known to exhibit antiferromagnetic (AFM) behavior, one may anticipate pursuing an A-type AFM configuration, where the Fe$^{3+}$ spins are aligned parallel within the same crystallographic layer but antiparallel between adjacent layers, corresponding to a layered AFM structure. This magnetic arrangement is illustrated in Fig.~\ref{fig:crystal}-A. However, self-consistent field (SCF) calculations for this configuration yielded a nonzero total magnetization (Fig.~\ref{fig:magnetization}-A), indicating that the A-type AFM ordering does not correctly capture the true ground-state magnetic structure of CuFeO$_2$. Instead, following the approach described in Ref.~\onlinecite{Ferri2019ThermodynamicEnvironments}, we adopted a G-type antiferromagnetic configuration, where the Fe spins alternate in all nearest-neighbor directions, resulting in complete cancellation of spin moments even within each layer.

\begin{figure}[!b]
\vspace{-6pt}
\begin{center} 
    \resizebox{.99\textwidth}{!}{\includegraphics{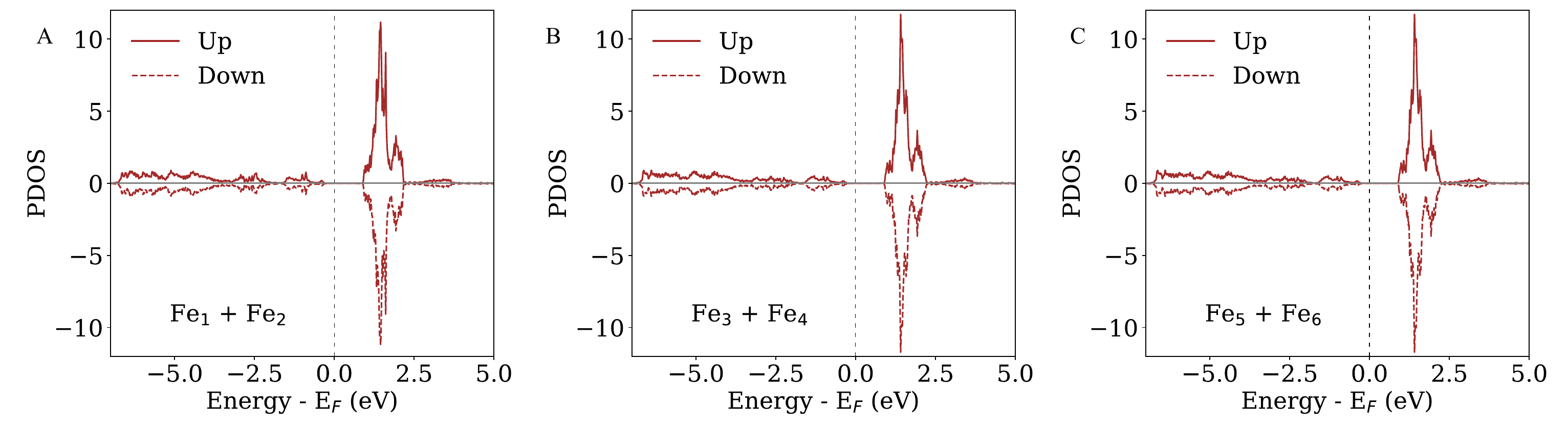}}

\end{center}
\vspace{-24pt}
\caption{\label{fig:fe-pdos} Fe-projected magnetization of each layer in CuFeO$_2$ supercell. (A): magnetization of the top layer, (B): magnetization of the middle layer, and (C): magnetization of the bottom layer.}
\vspace{-4pt}
\end{figure}

To accommodate this spin arrangement, we constructed a $2 \times 1 \times 1$ supercell, containing 24 atoms in total: 6 Fe, 6 Cu, and 12 O atoms. The spin configuration for each Fe atom in this G-type AFM structure is shown in Fig.~\ref{fig:crystal}-B. The SCF results for this configuration, shown in Fig.~\ref{fig:magnetization}-B, confirm that the total magnetization converges to zero---an essential criterion for correctly modeling an antiferromagnetic material. Figure~\ref{fig:fe-pdos} also confirms the G-type AFM behavior of CuFeO$_2$ as we can see the magnetization cancels within each layer.

Having thus satisfied the antiferromagnetic nature of CuFeO$_2$, we performed all subsequent electronic structure calculations using the $2 \times 1 \times 1$ supercell. To accurately describe the localized nature of the Fe 3d electrons, we applied a Hubbard $U$ correction, following widely adopted practices in transition metal oxide studies~\cite{Naveas2023First-principlesDFT+U+V}. We computed the Hubbard $U$ parameter using a recently developed linear-response approach within the framework of density-functional perturbation theory (DFPT)~\cite{Timrov2022HPTheory}. Specifically, we obtained $U$ values for all magnetically distinct Fe atoms from a single-shot DFPT iteration, and consistently used these values for both the DOS and band structure calculations. Table.~\ref{table1} summarizes these $U$ values, which we calculated using both atomic projection and orthogonalized atomic projection schemes~\cite{Cococcioni2005LinearMethod}. We performed all the DFT calculations using the QUANTUM ESPRESSO package~\cite{Giannozzi2009QUANTUMMaterials, Giannozzi2017AdvancedESPRESSO, Giannozzi2020QExascale}. We use a 90 Rydberg kinetic energy cutoff for the wavefunction, and a 1150 Rydberg kinetic energy cutoff for the charge density. For the SCF calculations, we sampled the Brillouin zone with a Monkhorst–Pack $6\times12\times2$ k-point grid, followed by a denser $12\times24\times4$ grid for non-self-consistent (NSCF) runs. We evaluated the electronic band structure along high-symmetry paths using 133 k-points.

\begin{figure}[!b]
\vspace{-6pt}
\begin{center} 
    \resizebox{.63\textwidth}{!}{\includegraphics{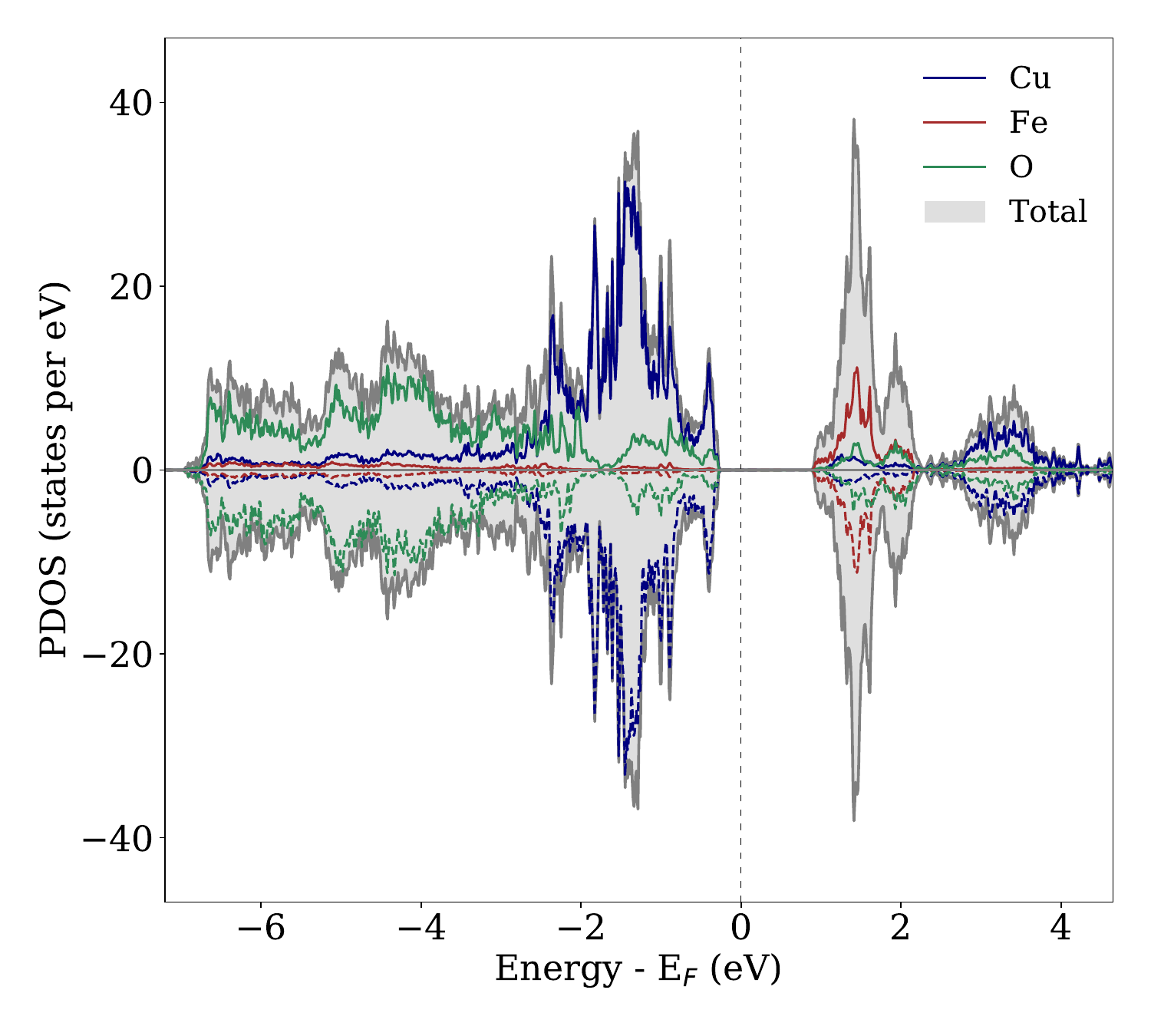}}
 
\end{center}
\vspace{-24pt}
\caption{\label{fig:dos} Projected and total density of States in CuFeO$_2$ using orthogonalized atomic projection for Hubbard U parameter. }
\vspace{-4pt}
\end{figure}

\begin{table}[!b]
    \centering
    \vspace{-1.5pt}
    \hspace{-3pt}
    \begin{tabular}{|l||c|c|} 
    \hline
    Fe & atomic & ortho-atomic \\ 
     \hline
     \hline
    Fe$_{1}$   & 4.3960 & 5.4638 \\
    Fe$_{2}$   & 4.3960 & 5.4638 \\
    Fe$_{3}$   & 4.4022 & 5.4680 \\
    Fe$_{4}$   & 4.4022 & 5.4680 \\
    Fe$_{5}$   & 4.4022 & 5.4680 \\
    Fe$_{6}$   & 4.4022 & 5.4680 \\ \hline
    \end{tabular}
    
    \caption{\label{table1} Computed Hubbard $U$ parameters for 6 Fe centers using atomic and ortho-atomic projection type.}
    \vspace{-4pt}
\end{table}

Since we performed the band structure calculations using a supercell, we applied the following band unfolding procedure to recover the band structure for the primitive cell:
\begin{itemize}
    \item Generated the high-symmetry k-path for the primitive cell using the VASPKIT tool~\cite{Wang2021VASPKIT:Code}.
    \item Constructed the corresponding k-path in the supercell using the open-source Python package BandUPpy~\cite{Medeiros2014EffectsUnfolding, Medeiros2015UnfoldingOperator, Iraola2022IrRep:Structures}.
    \item Performed the SCF and NSCF (band structure) calculations in the supercell using the generated k-points.
    \item Unfolded the resulting band structure using BandUPpy to obtain the band structure in the Brillouin zone of the primitive cell. The unfolded bands come with weights that indicate the extent to which each supercell Bloch state contributes to the corresponding primitive cell state.
\end{itemize}

\begin{figure}[!h]
\vspace{-6pt}
\begin{center} 
    \resizebox{.7\textwidth}{!}{\includegraphics{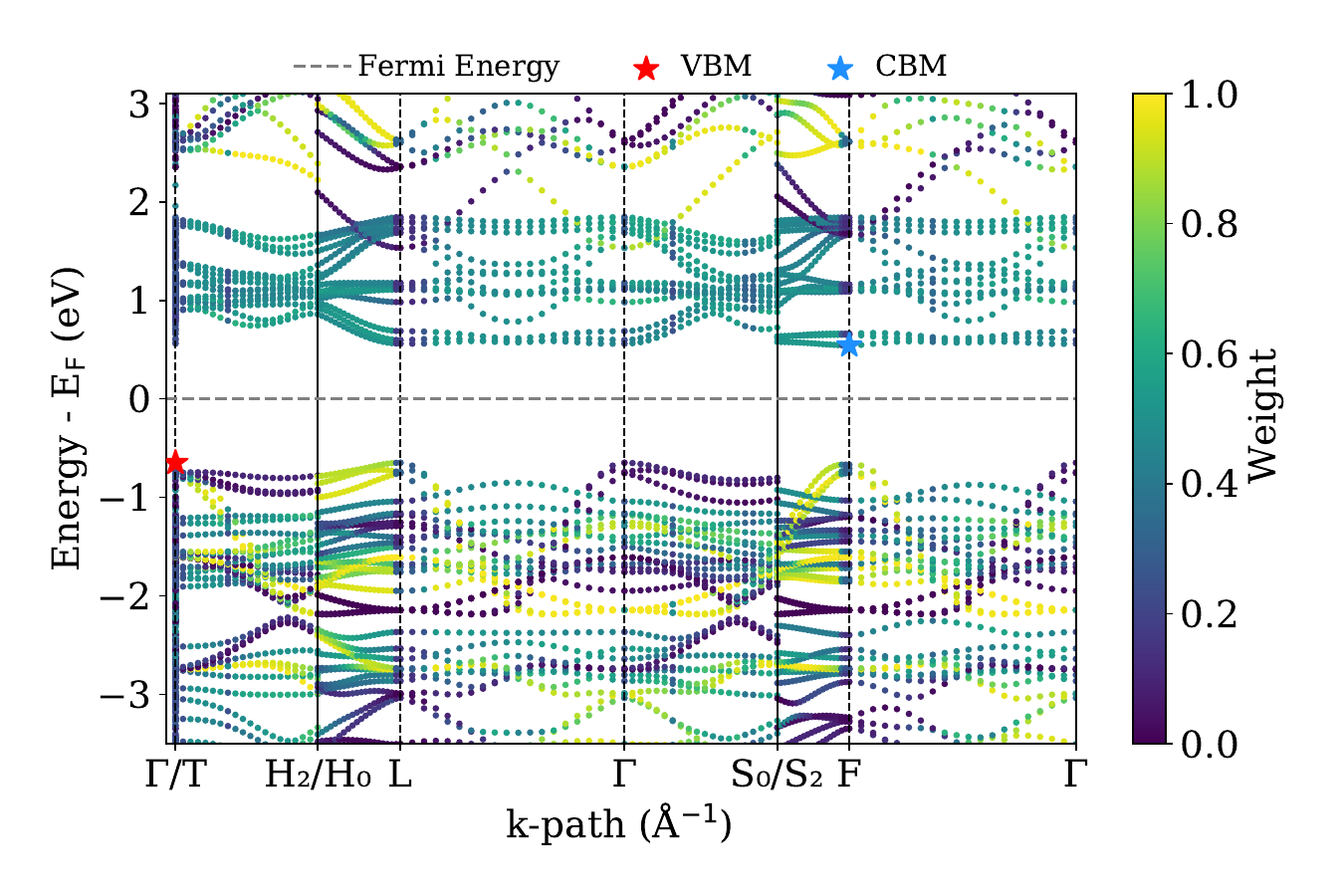}}
\end{center}
\vspace{-24pt}
\caption{\label{fig:band} Band structure of CuFeO$_2$ using orthogonalized atomic projection for Hubbard U parameter. }
\vspace{-4pt}
\end{figure}

In the main text, we show the projected DOS and unfolded band structure using Hubbard $U$ parameters obtained via atomic projection. Here, we also report the projected DOS and unfolded band structure with $U$ parameters using the orthogonalized atomic projection scheme. Figures~\ref{fig:dos} and~\ref{fig:band} show the DOS and band structure obtained when using orthogonalized atomic projection for the determination of the Hubbard $U$. We find that the bandgap in CuFeO$_2$ is approximately 0.95~eV when using atomic projection and 1.19~eV with the orthogonalized atomic projection scheme. These values are consistent with previous literature values that estimate the bandgap between 0.8-1.5~eV~\cite{Xu2021RevealingSemiconductor, Maouhoubi2024StructuralCompound, Upadhyay2020ComputationalCuFO2}.

\section{Theoretical Calculations of Ground and Excited State Core-Level Spectra}

The theoretical framework for interpreting the experimental transient core-level spectra has been described previously~\cite{Klein2022iAb/isub3/sub}. In brief, we employ an \textit{ab initio} combined theoretical approach based on density functional theory (DFT) and the Bethe--Salpeter equation (BSE). This approach is implemented using the \textsc{Quantum ESPRESSO} package~\cite{Giannozzi2009QUANTUMMaterials, Giannozzi2017AdvancedESPRESSO, Giannozzi2020QExascale} along with modifications to the existing OCEAN code (Obtaining Core-level Excitations using \textit{Ab Initio} Calculations and the NIST BSE solver)~\cite{Vinson2011Bethe-SalpeterSpectra, Vinson2022AdvancesPackage, Gilmore2015EfficientCalculations, Shirley2020TheExcitations}. These modifications enable the inclusion of excited-state distributions, allowing us to determine how the ground-state band structure relates to the observed transient changes in the XUV spectra.

\subsection{Fe $\rm M_{2,3}$ Edge Ground State Calculations}
The OCEAN package does not support projector-augmented wave (PAW) pseudopotentials, so the DFT portion of our BSE calculations was conducted using Troullier--Martins norm-conserving pseudopotentials within the generalized gradient approximation (GGA) and the Perdew--Burke--Ernzerhof (PBE) functional~\cite{Perdew1996GeneralizedSimple, Troullier1991EfficientCalculations}. Prior to the OCEAN calculations, a $2 \times 2 \times 1$ supercell of antiferromagnetic (AFM) ordered CuFeO\textsubscript{2} in a rhombohedral geometry (3R-CuFeO\textsubscript{2}) was structurally relaxed using the \textsc{Quantum ESPRESSO} package~\cite{Giannozzi2009QUANTUMMaterials, Giannozzi2017AdvancedESPRESSO, Giannozzi2020QExascale}. The lattice constants for the primitive unit cell were $a = b = 3.0349$~\AA{} and $c = 17.1656$~\AA~\cite{Jiang2019ElectronicPhotocathodes}. A Hubbard $U$ value of 2~eV, a 300~Rydberg energy cutoff, and a $3 \times 3 \times 1$ \textit{k}-point grid were employed for the structural relaxation. The optimized geometry was used in the BSE calculations of the XUV ground-state reflectivity at the Fe M\textsubscript{2,3} X-ray edge using the OCEAN code~\cite{Vinson2011Bethe-SalpeterSpectra, Vinson2022AdvancesPackage, Gilmore2015EfficientCalculations, Shirley2020TheExcitations}. The BSE equations were solved using a $6 \times 6 \times 6$ \textit{k}-point mesh, a dielectric constant of 21.8, a 0.7 scaling factor for the Slater $G$ parameter, and a 300~Rydberg energy cutoff. A 4.0~Bohr screening radius and a $2 \times 2 \times 2$ screening mesh were applied. A post-calculation broadening routine was implemented using a MATLAB script to match the linewidth broadening observed in the experiment.

\subsection{Fe $\rm M_{2,3}$ Edge Excited State Calculations}

Excited state changes to the transient XUV spectra at the Fe M\textsubscript{2,3} edge of CuFeO\textsubscript{2} were modeled using a modification to the OCEAN package discussed previously~\cite{Liu2023MeasuringSpectroscopy, Klein2022iAb/isub3/sub, Vinson2022AdvancesPackage, Shirley2020TheExcitations}. The modifications to the OCEAN code enables state-blocking of core-to-valence X-ray transitions at different points in momentum space, which simulates state-filling of the valence and conduction bands following 400~nm photoexcitation. The state-blocked population of electrons and holes forbids or allows XUV transitions to specific points in \textit{k}-space but does not account for carrier density, therefore, we use the peak position and relative intensity of the modeled spectra to compare experiment to theory. We obtain the modeled excited-state XUV spectra by subtracting the calculated ground state spectrum from the calculated excited state spectrum to calculate the differential absorption. Figure~3A in the main text demonstrates agreement between the BSE modeled 400~nm charge transfer state and experiment.

\begin{figure}[!t]
\vspace{-6pt}
\begin{center} 
    \resizebox{.54\textwidth}{!}{\includegraphics{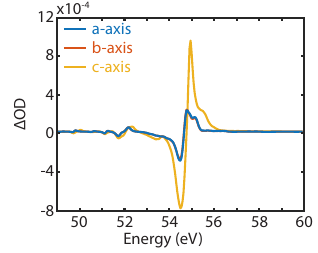}}

\end{center}
\vspace{-24pt}
\caption{\label{fig:s4} DFT+BSE calculated lattice expansion and polaron distortion for an anisotropic \textit{a}-axis (blue), \textit{b}-axis (red), or \textit{c}-axis (yellow) expansion
 }
\vspace{-4pt}
\end{figure}

We semi-empirically model the transient effects of distortions caused by the polaron state and the anisotropic \textit{c}-axis expansion using a framework described previously~\cite{Klein2022iAb/isub3/sub}. We model polarons using the bond distortion method in which we locally apply an expansion to Fe--O bonds at one iron octahedra in the CuFeO\textsubscript{2} supercell~\cite{Pham2020EfficientMethods}. The anisotropic, local distortion of the polaron was modeled as an expansion of the iron octahedra bond lengths ranging from 2--8\% and then applied to the OCEAN code to calculate the resulting core-valence spectrum. The resulting modeled differential spectrum plotted in Figure~3 of the main text is from a 6\% expansion of the local iron octahedra selected for polaron formation. We also model an overall 6\% anisotropic lattice expansion on each of the lattice directions to establish which lattice direction is responsible for the signal we observe after 2~ps in our transient spectrum. When modeling the overall anisotropic lattice expansion, we selectively expanded the lattice constants based upon the chosen axis to model which best agreed with experiment. Figure~\ref{fig:s4} shows a plot of \textit{a}-, \textit{b}-, and \textit{c}-axis expansions.

We find that the \textit{c}-axis expansion best agrees with experiment. Particularly, the \textit{c}-axis expansion is blue-shifted like the polaron feature, suggesting that expansion along the \textit{c}-axis contributes most to the convolved lattice expansion and polaron signal that we observe in our experimental transient XUV spectrum. The \textit{c}-axis expansion has a much stronger contribution to the spectral signal than expansions along the other two axes. When convolved with the polaron feature, the \textit{a}- and \textit{b}-axis expansions would not result in the same spectral line shape.

\section{Transient XUV Spectral Analysis}

\begin{figure}[!t]
\vspace{-6pt}
\begin{center} 
    \resizebox{.54\textwidth}{!}{\includegraphics{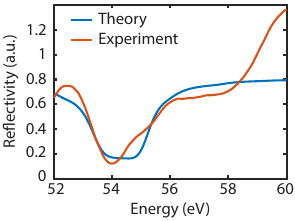}}
\end{center}
\vspace{-24pt}
\caption{\label{fig:s5} Static XUV reflectivity (orange) and BSE calculated ground state (blue) spectra of CuFeO\textsubscript{2} thin film.}
\vspace{-4pt}
\end{figure}

\begin{figure}[!b]
\vspace{-6pt}
\begin{center} 
    \resizebox{.54\textwidth}{!}{\includegraphics{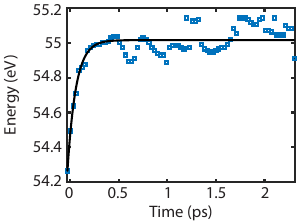}}
\end{center}
\vspace{-24pt}
\caption{\label{fig:s6} Exponential fit of the spectral shift at the Fe $\rm M_{2,3}$ edge of CuFeO\textsubscript{2}.}
\vspace{-4pt}
\end{figure}

The ground state reflectivity of the CuFeO\textsubscript{2} thin film around the Fe M\textsubscript{2,3} edge is shown in Fig.~\ref{fig:s5}. The static XUV reflectivity of CuFeO\textsubscript{2} is obtained by normalizing the measured static reflectivity spectrum of CuFeO\textsubscript{2} to the static reflectivity of a Si wafer, which does not have any absorption features below 100~eV. There is good agreement between the experimental ground state XUV reflectivity of CuFeO\textsubscript{2}, and our BSE calculated ground state (Fig.\ref{fig:s5}).

Figure~\ref{fig:s6} shows an exponential fit of the spectral blue shift associated with photoexcited polaron formation at the Fe M\textsubscript{2,3} edge was performed. A single exponential function following the equation $a \cdot \exp(-x/b) + c$ results in a time constant of $90 \pm 20$~fs. The spectral shift was measured to have a magnitude of $0.7 \pm 0.1$~eV.

\begin{figure}[!t]
\vspace{-6pt}
\begin{center} 
    \resizebox{.54\textwidth}{!}{\includegraphics{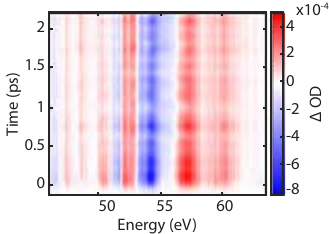}}
\end{center}
\vspace{-24pt}
\caption{\label{fig:s7} Reconstructed transient spectrum based upon the singular value decomposition of the transient XUV at the Fe $\rm M_{2,3}$ edge. The vectors employed for the SVD are shown in Figure 3A corresponding to the charge transfer state (blue, left), the polaron state (green, middle), and the lattice expansion and polaron state (red, right).}
\vspace{-4pt}
\end{figure}

Singular value decomposition (SVD) of the transient spectrum was performed to confirm independent spectral features and their kinetics and to ascribe weights to each independent spectral feature with respect to time. In the SVD analysis, we factorize the two-dimensional matrix of the transient spectrum by rotating (U), scaling (S), and then again rotating (V) the matrix. The scaling matrix contains the singular values along its diagonal, which we refer to as the number of components. This results in 3 singular values being necessary to adequately reconstruct the transient XUV spectrum (Fig.~\ref{fig:s7}. Each singular value corresponds to the three kinetic states that we observe in our transient spectrum. We ascribe the first singular value to the charge transfer state at 30~fs, the second to the polaron state at 180~fs, and the third to the convolved polaron and \textit{c}-axis lattice expansion at 2~ps. We find that the inclusion of three singular values in our decomposition is adequate to reconstruct our transient spectra.

\begin{figure}[!h]
\vspace{-6pt}
\begin{center} 
    \resizebox{.97\textwidth}{!}{\includegraphics{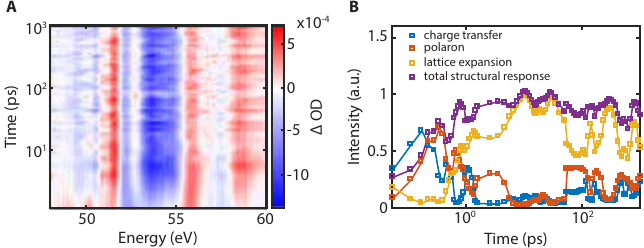}}
\end{center}
\vspace{-24pt}
\caption{\label{fig:s8} Reconstructed transient spectrum based upon the singular value decomposition of the transient XUV at the Fe $\rm M_{2,3}$ edge. The vectors employed for the SVD are shown in Figure 3A corresponding to the charge transfer state (blue, left), the polaron state (green, middle), and the lattice expansion and polaron state (red, right).}
\vspace{-4pt}
\end{figure}

To confirm that three singular values are necessary to accurately describe the kinetic processes that take place in CuFeO\textsubscript{2} following 400~nm photoexcitation, we also measure the transient XUV response out to 1~ns. The long timescale measurements of thin film CuFeO\textsubscript{2} can be seen in Fig.~\ref{fig:s8}A and do not experience any additional phenomena out to 1~ns. Further, the long timescale measurement demonstrates that the convolved polaronic and lattice expansion feature lives out to nanoseconds. We also include a kinetic plot of the contributions of the singular value decomposition components out to 1~ns (Fig.~\ref{fig:s8}B).

\section{Quantum dynamics simulation of polaron formation}

The Holstein Hamiltonian~\cite{Holstein1959StudiesMotion} has been widely invoked to understand polaron formation and transport in materials ranging from organic crystals to transition metal dichalcogenides~\cite{Ortmann2009TheoryModel, Bhattacharyya2025NonequilibriumDiffusion}. While the traditional Holstein Hamiltonian considers interactions between electronic excitations (charge carriers or excitons) and single phonon mode on each lattice site, we adopt the dispersive Holstein model~\cite{Bhattacharyya2024AnomalousBoundaries}, which couples a carrier on a given site to phonons spanning a distribution of phonon frequencies, consistent with the variety of phonon modes observed in Raman spectra of materials. The dispersive Holstein Hamiltonian takes the form 
\begin{equation}
    \hat{H} = \sum_{i}^{N}\epsilon_i \hat{n}_i + \sum_{\langle ij \rangle }^{N} v_{ij} \hat{a}_i^\dag \hat{a}_j + \frac{1}{2}\sum_{\alpha} \big[ \hat{P}_{i,\alpha}^2 +  \omega_{i\alpha}^2 \hat{X}_{i,\alpha}^2 \big] + \sum_{\alpha} c_{i, \alpha} \hat{X}_{i,\alpha} \hat{n}_i \label{eq:ham}.
\end{equation}
Here, $\hat{a}_i$ ($\hat{a}_i^\dag$) is the annihilation (creation) operator for a charge carrier at lattice site $i$, and $\hat{n}_i = \hat{a}_i^\dag \hat{a}_i$. $\hat{X}_{i,\alpha}$ ($\hat{P}_{i,\alpha}$) is the dimensionless position (momentum) operator for the $\alpha$-th phonon mode connected with the $i$-th lattice site. The first term in the Hamiltonian (Eq.~\ref{eq:ham}) represents the on-site energy of the carrier, and the second term is responsible for carrier hopping. The angular brackets $\langle \cdots \rangle$ guarantee the nearest-neighbor hopping. We take all lattice sites to be degenerate ($\epsilon_i = 0$), and assume a uniform site-hopping parameter ($v_{ij} = v = 300$~cm$^{-1}$) consistent with previous studies on calculations of the hopping integral $v$ in hematite and other iron-containing transition metal oxides~\cite{Ahart2020PolaronicOxides, Ahart2022ElectronTheory}. The third term in Eq.~\ref{eq:ham} consitutes the site-specific phonon baths, and the last term indicates on-site carrier-phonon coupling, which is linear in the local phonon bath coordinates and is responsible for the fluctuation of the on-site energies, ultimately resulting in small polaron formation. 

We describe the carrier-phonon interaction with a spectral density 
\begin{equation}\label{spec_dens_def1}
   J_i(\omega)= \dfrac{\pi }{2} \sum_{\alpha} \dfrac{c_{i,\alpha}^2} {\omega_{i\alpha}}\delta(\omega -\omega_{i\alpha}),
\end{equation}
choose these to be equivalent across all sites, and employ a sum of Lorentzians to model their distributions 
\begin{equation}\label{eq:lorentz}
    J(\omega) = \sum_{i=1}^{N_m} 4 \lambda_i \frac{\Omega_i^2 \gamma_j \omega}{(\omega^2 - \Omega_i^2)^2 + 4 \gamma_i^2 \omega^2}. 
\end{equation}
Here, $N_m$ denotes the number of Lorentzian distributions (also known as Brownian oscillator densities) we use in our model, $\Omega_i$ and $\gamma_i$ denote the position of the peak and the width of the $i$-th Lorentzian distribution, and the coefficient $\lambda_i$ scales the spectral density to the correct reorganization energy $\lambda$. For any spectral density $J(\omega)$, the total reorganization energy is defined as 
\begin{equation}
    \lambda = \frac{1}{\pi} \int d\omega \frac{J(\omega)}{\omega} = \sum_i \lambda_i.
\end{equation}
To parameterize the spectral densities for CuFeO$_{2}$, we adopt values obtained in previous calculations and inferred from experiment. Specifically, we select the Lorentzian parameters (peak height and width) to align with clusters of experimentally observed Raman peaks in CuFeO$_2$~\cite{Chen2021Temperature-dependentTransition}. For the reorganization energies, we consider previous work on $\alpha$-Fe$_2$O$_3$, which has shown that $\lambda / v \approx 9$, with $\lambda_{\rm Fe_2O_{3}} \approx 2984$~cm$^{-1}$~\cite{Ahart2020PolaronicOxides}. In contrast, experimental evidence in CuFeO$_{2}$ suggests a possible reduction in the reorganization energy. To reflect this, we adopt a reduced ratio of $\lambda/v \approx 5$ for CuFeO$_{2}$. Hence, $\lambda_{\rm CuFeO_{2}} \approx 1500$~cm$^{-1}$. Table~\ref{table2} summarizes the parameters we employ in this study.

 \begin{table}[!b]
    \centering
    \vspace{-1.5pt}
    \hspace{-3pt}
    \begin{tabular}{|l||c|c|c|c|} 
    \hline
    Region & $\Omega_i$ [cm$^{-1}$]  & $\lambda_i$ [cm$^{-1}$] & $\gamma_i$ [cm$^{-1}$] & Raman peaks [cm$^{-1}$] \\ 
     \hline
     \hline
    Low-frequency region   & 545 & 1500 & 180 & 352, 509 \& 652\\ \hline
    High-frequency region   & 1100 & 1500 & 100 & 1000, 1050 \& 1171\\ \hline
    Both regions   & 545 \& 1100 & 900 \& 600 & 180  \& 100 & 352, 509, 652, 1000, 1050 \& 1171\\ \hline
    \end{tabular}
    
    \caption{\label{table2} Parameters for Lorentzian spectral densities in Eq.~\ref{eq:lorentz}. For all three cases, we fix the total reorganization energy $\lambda = 1500$ cm$^{-1}$.}
    \vspace{-4pt}
\end{table}

We simulate the quantum dynamics of polaron formation by invoking the \emph{Non‑Interacting Blip Approximation (NIBA)}, a second-order perturbative theory valid when $v / \lambda \ll 1$~\cite{Dekker1987Noninteracting-blipBath}. This condition holds in our system. Under NIBA, the carrier population $P_n(t)$ of each site evolves according to a simple integrodifferential equation:
\begin{equation}\label{eq:niba}
    \frac{dP_n(t)}{dt} = \sum_{m=1}^N \int_0^t \,d\tau \mathcal{K}_{nm} (t, \tau) P_m (\tau).
\end{equation}
Here, the kernel $\mathcal{K}(t)$ arises from a second order expansion in the hopping integral $v$, with non‑perturbative, multiphonon treatment of the carrier-phonon coupling. With $\epsilon_i = 0$, the elements of the kernel, $\mathcal{K}(t)$, simplify thus,
\begin{equation}
    \mathcal{K}_{n \neq m}(t, \tau) = 2 |v_{nm}|^2\exp[-Q_2 (t-\tau)] \cos\Big[Q_1(t-\tau)- (1+\frac{\delta_n + \delta_m}{2})(Q_1(t)-Q_1(\tau))\Big],
\end{equation}
and, $\mathcal{K}_{nn}(t, \tau) = -\sum_{m \neq n} \mathcal{K}_{mn}(t, \tau)$. The bath correlation function $Q(t) = Q_2 (t) + iQ_1(t)$, is defined as 
\begin{equation}
    Q(t) = 2\pi \int_{0}^{\infty} \,d\omega\frac{J(\omega)}{\omega^{2}} \, \left\{ \coth\!\bigl(\tfrac{\beta\omega}{2}\bigr)\,[1 - \cos(\omega t)] \;+\; i\,\sin(\omega t) \right\}.
\end{equation}
The shift parameter $\delta$ arises from the initial bath conditions. For simulations starting from an \textit{unpolarized} initial condition, $\delta = 0$ for all sites, whereas, for those with \textit{polarized initial conditions}, $\delta = -1$ for the initial site and  $\delta = 0$ for all other sites. 

\begin{figure}[!t]
\vspace{-6pt}
\begin{center} 
    \resizebox{.95\textwidth}{!}{\includegraphics{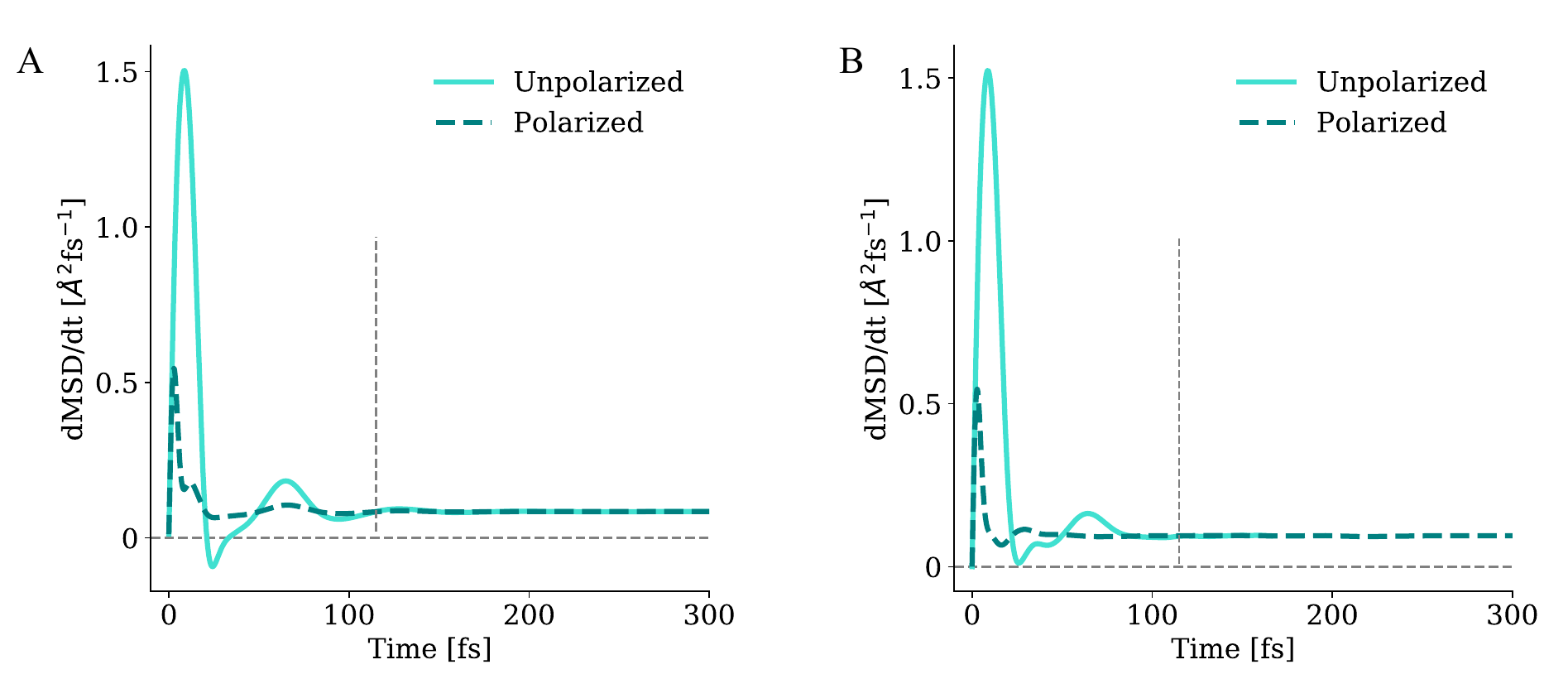}}
\end{center}
\vspace{-24pt}
\caption{\label{fig:dynamics} Derivative of the mean squared displacement of polarons in the dispersive Holstein lattice obtained using (A) the numerically exact HEOM and (B) the perturbative NIBA. Both methods show qualitative agreement and, crucially, predict similar polaron formation timescales (within $\pm 15$~fs error).}
\vspace{-4pt}
\end{figure}

We validate our NIBA-based dynamics by comparing against numerically exact results obtained using the Hierarchical Equations of Motion (HEOM)~\cite{Tanimura1989TimeBath}. Because HEOM becomes prohibitively computationally expensive for this strong carrier-phonon coupling limit when using multi-peaked spectral densities, we perform this comparison using a single-peak Lorentzian spectral density with parameters equivalent to those of the low-frequency component in Table~\ref{table2}. Following Ref.~\onlinecite{Li2022ADensity}, we performed HEOM simulations for both polarized and unpolarized initial conditions. Figure~\ref{fig:dynamics}-A displays the HEOM results, while panel (B) shows the corresponding NIBA dynamics. Although NIBA captures the quantum dynamics for both initial conditions only semiqualitatively, it predicts the correct polaron formation timescale, i.e., the timescale corresponding to the place where the nonequilibrium dynamics arising from these initial conditions start to agree. The vertical dashed lines in both panels of Fig.~\ref{fig:dynamics} indicate the polaron formation timescale. In summary, although NIBA is an approximate, perturbative theory, it accurately captures polaron formation timescales compared against numerically exact methods.

To lower the cost of our quantum dynamics simulations and ensure that we can reach the thermodynamic limit of polaron formation free of finite-size effects, we adopt our space-time local generalized master equation (STL-GME) framework \cite{Bhattacharyya2025spacelocal}. The STL-GME recognizes that systems like small polaron formers have finite spatial memory. We exploit this short-ranged memory to enable more efficient simulation of the quantum dynamics. This only entails moving from the time-nonlocal formulation in Eq.~\ref{eq:niba} to a time-local and integrated one:
\begin{equation}
    C_{n,m}(t+\delta t) = \sum_k \mathcal{U}_{n,k}(t) C_{k,m}(t),
\end{equation}
where $\mathcal{U}$ is the dynamical generator for the spread of population dynamics. This quantity equilibrates to a finite value at short times, over which time a polaron can only spread over a finite spatial extent. As such, its form in an even larger space remains the same, enabling us to perform the quantum dynamics simulation of polaron formation over a large space using only a small reference calculation over a smaller space. In our current calculations, adopting the STL-GME reduces the memory cost by about an order of magnitude: a 15-site NIBA calculation with a 330~fs simulation time requires $\sim$80~GB of memory, while the STL-GME reduces the cost to that of a 9-site calculation with a $\sim$200~fs simulation time, requiring only about $\sim 10$~GB of memory. We achieve convergence between NIBA and STL-GME in the derivative of the MSD, with an average error of $\sim 1.5\%$. For all MSD calculations, we employ a lattice distance of 5~\AA.

We conclude our discussion with a short analysis of the conditions on the spectral density that determine relative speeds of polaron formation. In particular, we interrogate the seemingly counterintuitive result that adding a \textit{fast} \textit{Lorentzian} component to the spectral density in our model of CuFeO$_2$ \textit{slows} polaron formation---a finding that contrasts earlier studies demonstrating that a \textit{faster} \textit{Debye} bath correlates to \textit{faster} polaron formation~\cite{Bhattacharyya2024AnomalousBoundaries}. We argue that this apparent contradiction is an artifact of the Debye spectral density, which conflates the characteristic frequency of the coupling distribution (its maximum) with its width into one parameter, $\omega_c$. Instead, the Lorentzian spectral density disentangles the characteristic frequency of the coupling distribution, $\Omega$, from the width of the distribution, $\gamma$, allowing us to test the effect of characteristic bath speed separately from the breadth of the phonon energies that are accessible.

\begin{figure}[!t]
\vspace{-6pt}
\begin{center} 
    \resizebox{.6\textwidth}{!}{\includegraphics{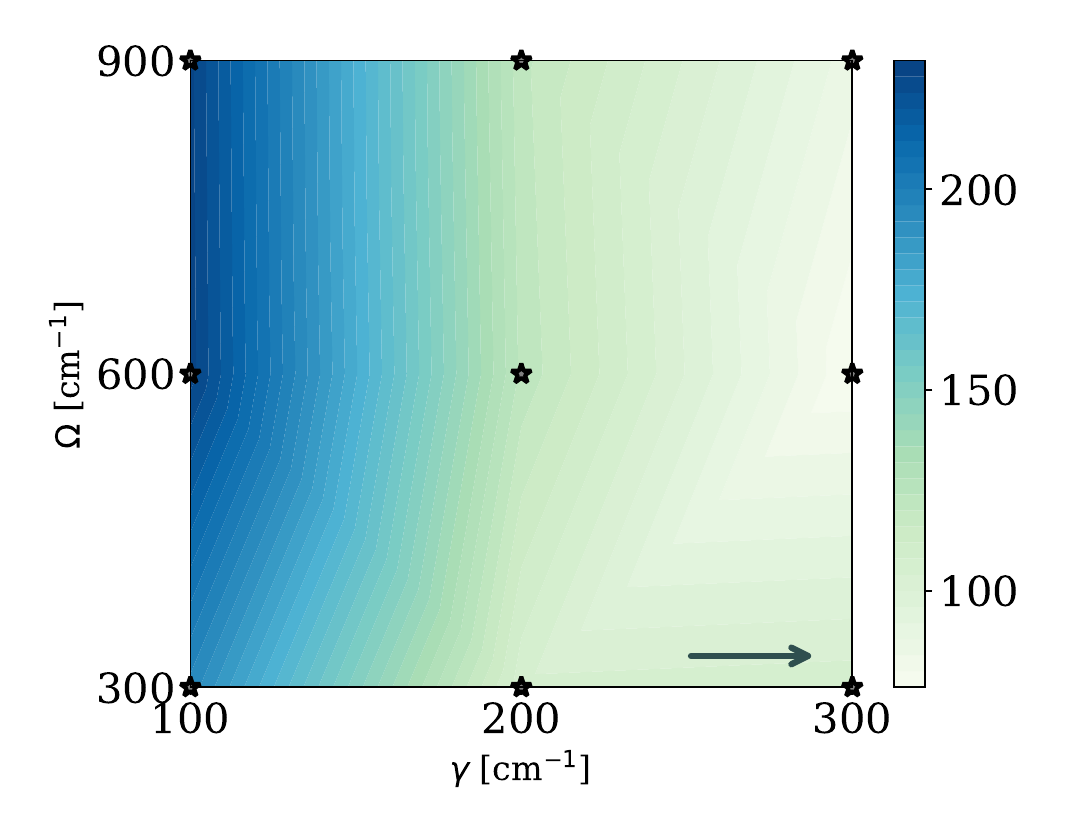}}
\end{center}
\vspace{-24pt}
\caption{\label{fig:ph_space} Polaron formation time as a function of the Lorentzian peak position, $\Omega$, and peak width, $\gamma$. The arrow shows that the decrease in polaron formation times observed at fixed $\Omega$ as $\gamma$ increases mirrors the decrease in polaron formation times observed with increasing characteristic frequency, $\omega_c$, in the Debye spectral density, consistent with our previous results detailed in Ref.~\onlinecite{Bhattacharyya2024AnomalousBoundaries}.}
\vspace{-4pt}
\end{figure}

To demonstrate this decoupling argument in the Lorentzian spectral density, we analyze the dynamics arising from a wide range of Lorentzian distributions by varying both $\Omega$ and $\gamma$. Figure~\ref{fig:ph_space} reports the polaron formation timescale as a function of $\Omega$ and $\gamma$. We find that these timescales are highly sensitive to the width, $\gamma$, but largely insensitive to the peak position, $\Omega$. \textit{Physically, this is because polaron formation occurs faster when the phonon energies span a broad range, enabling more distributed energy dissipation.} To understand the contrast between the Lorentzian and Debye spectral densities, we consider the overdamped limit of $\gamma >> \Omega$, where one can approximate the Lorentzian distribution as a Debye 
\begin{equation}
    J(\omega) \approx \frac{2 \lambda (2 \gamma)} {\omega^2 + (2 \gamma)^2},
\end{equation}
with $\omega_c \rightarrow 2\gamma$. For Debye spectral densities, $\omega_c$ is the characteristic phonon frequency that dictates both the peak position and the width of the peak, and hence the decay timescale of the exponential of ${\rm Re\ }Q (t) = Q_2(t)$, $\mathrm{Exp}[-Q_2(t)]$, which dictates the dissipative behavior of the NIBA kernel $\mathcal{K} (t)$. Indeed, under this overdamped condition, we find that increasing $\gamma$ makes the polaron formation timescale shorter (see Fig.~\ref{fig:ph_space}), showing that our present results are in agreement with our previous observations~\cite{Bhattacharyya2024AnomalousBoundaries}. 

\begin{figure}[!b]
\vspace{-6pt}
\begin{center} 
    \resizebox{.46\textwidth}{!}{\includegraphics{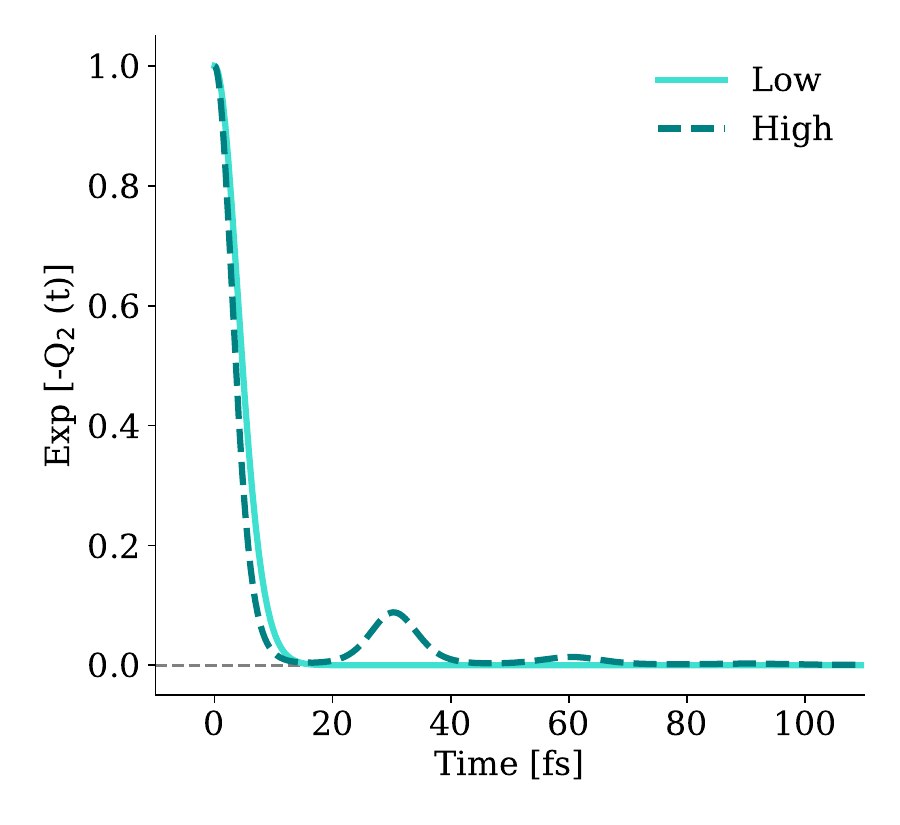}}
\end{center}
\vspace{-24pt}
\caption{\label{fig:qfunc} Comparison of the dissipative component of the NIBA kernel, $\mathrm{Exp}[-Q_2(t)]$, for the Lorentzian spectral density in the presence of low-frequency versus high-frequency modes.}
\vspace{-4pt}
\end{figure}

We are now in a position to illustrate how the addition of the \textit{fast but narrow} Lorentzian spectral density can yield slower polaron formation timescales than a \textit{slow but broad} spectral density. Specifically, we focus on the decay behavior of $\mathrm{Exp}[-Q_2(t)]$. For the case with only low-frequency phonon modes, $\mathrm{Exp}[-Q_2(t)]$ shows a smooth decay, while for the case with only high-frequency contributions, $\mathrm{Exp}[-Q_2(t)]$ initially shows a decay, but later shows a recurrence pattern, as evident in Fig.~\ref{fig:qfunc}. A similar behavior in terms of polaron formation timescales is evident in Fig.~\ref{fig:ph_space}, where increasing $\gamma$ correlates strongly with shorter polaron formation timescales.

\begin{table}[!t]
    \centering
    \vspace{-1.5pt}
    \hspace{-3pt}
    \begin{tabular}{|c||cc|cc|} 
    \hline
    \: \: Cases \: \: &  \multicolumn{2}{c|}{Low-frequency region}   &  \multicolumn{2}{c|}{High-frequency region}  \\ 
     
        & $\Omega$ [cm$^{-1}$] & $\gamma$ [cm$^{-1}$] & $\Omega$ [cm$^{-1}$] & $\gamma$ [cm$^{-1}$] \\ \hline
   Case 1   & 545 & 180 & 1100 & 100 \\ \hline
    Case 2   & 545 & 100 & 1100 & 180 \\ \hline
    Case 3   & 545 & 180 & 1100 & 180 \\ \hline
    Case 4   & 545 & 180 & 1400 & 100 \\ \hline
    \end{tabular} 
    \caption{\label{table4} Parameters for the doubly peaked Lorentzian spectral densities considered in the four cases discussed in the text.}
    \vspace{-4pt}
\end{table}

\begin{figure}[!b]
\vspace{-6pt}
\begin{center} 
    \resizebox{.8\textwidth}{!}{\includegraphics{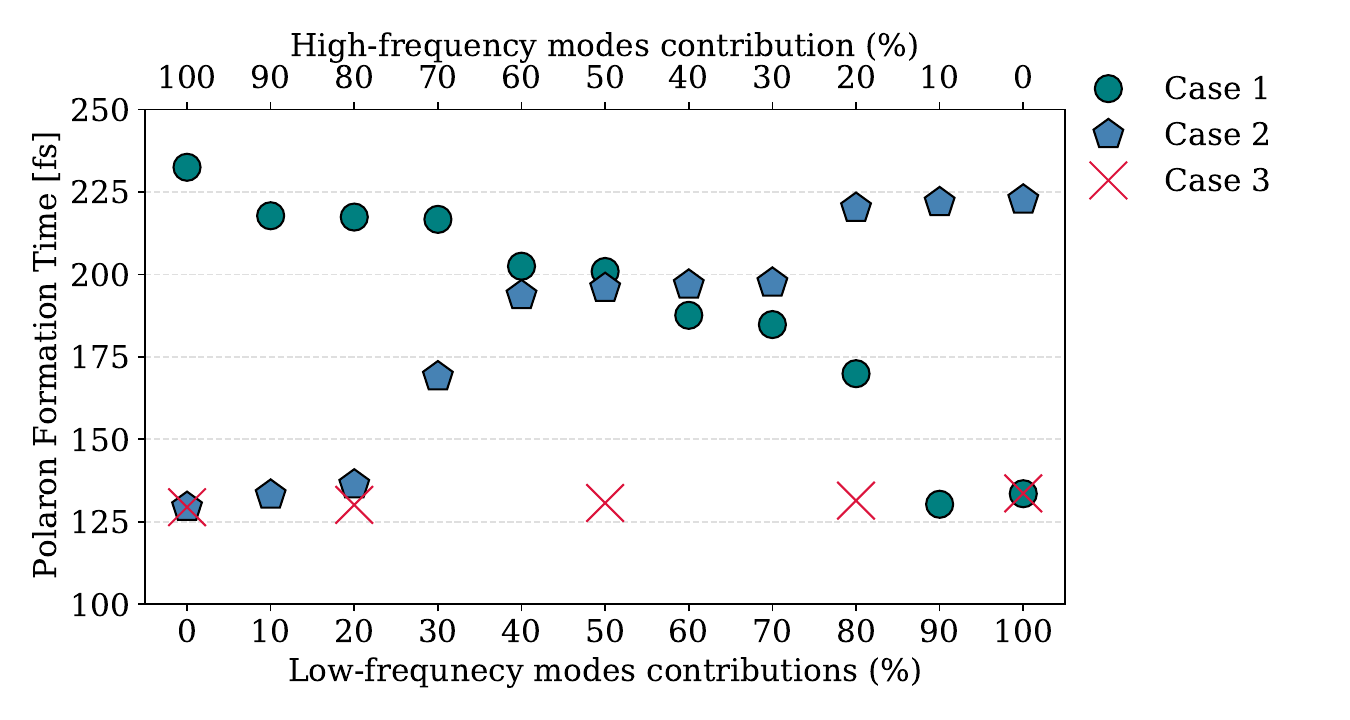}}
\end{center}
\vspace{-24pt}
\caption{\label{fig:comp1} Polaron formation times observed as we vary the balance of reorganization energies distributed between the low- and high-frequency regions for the different cases mentioned in Table~\ref{table4}.}
\vspace{-4pt}
\end{figure}

Having shown that the available phonon density is the main factor that dictates the polaron formation timescale for singly peaked Lorentzian densities, we turn to doubly peaked counterparts that are reminiscent of the division between acoustic and optical phonon modes in solid-state materials. In particular, we consider three paradigmatic cases specified in Table.~\ref{table4}. Case 1 refers to the distribution where the low-frequency peak has a broad phonon distribution and the high-frequency region has a narrow distribution. Case 2 corresponds to the opposite scenario with a broad high-frequency region and a narrow low-frequency one. In case 3, both low- and high-frequency regions have the same breadth. 

\begin{figure}[!b]
\vspace{-6pt}
\begin{center} 
    \resizebox{.8\textwidth}{!}{\includegraphics{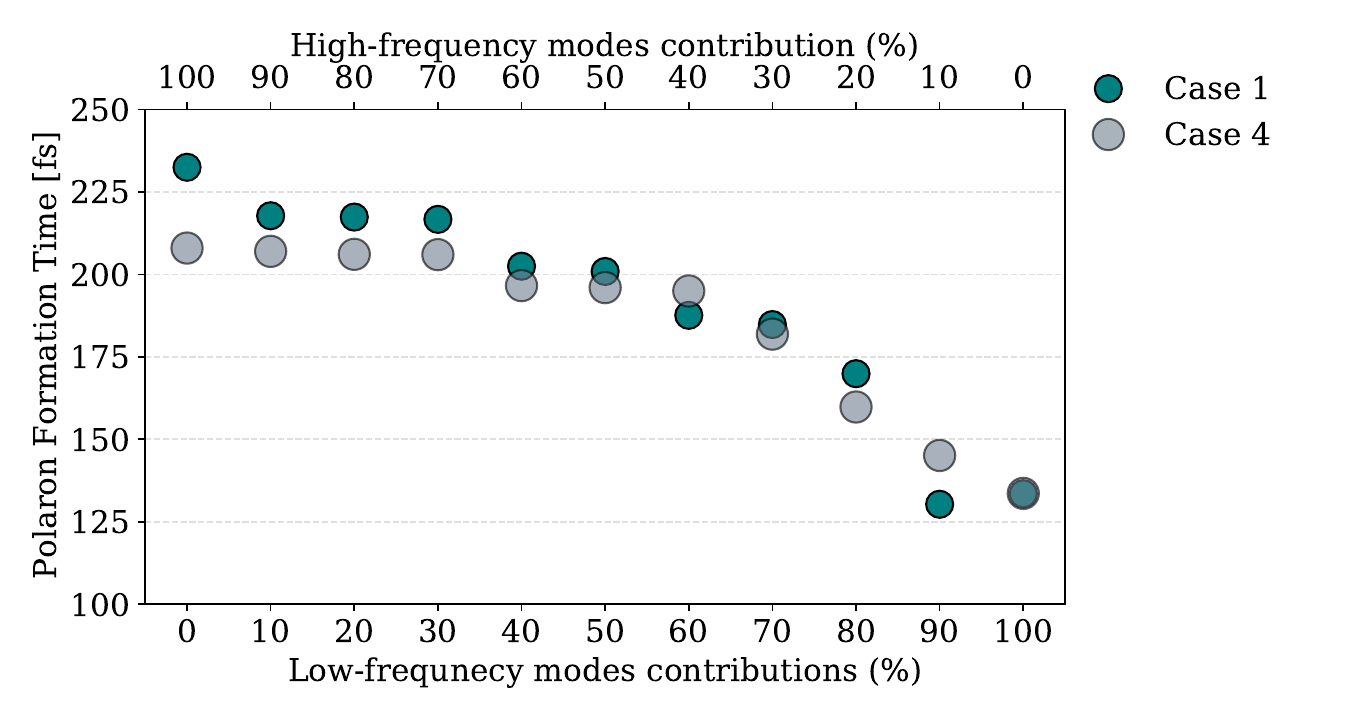}}
\end{center}
\vspace{-24pt}
\caption{\label{fig:comp2} 
Overall agreement in the polaron formation times observed across cases 1 and 4, mentioned in Table~\ref{table4}, as we vary the balance of reorganization energies distributed between their low- and high-frequency regions. Importantly, the only difference between cases 1 and 4 is the center frequency of the high- and low-frequency contributions. This agreement reiterates the insight that the factors that largely determine polaron formation timescales are the breadth of polaron distributions and their weights determined by their respective reorganization energies.}
\vspace{-4pt}
\end{figure}

\begin{figure}[!t]
\vspace{-6pt}
\begin{center} 
    \resizebox{.8\textwidth}{!}{\includegraphics{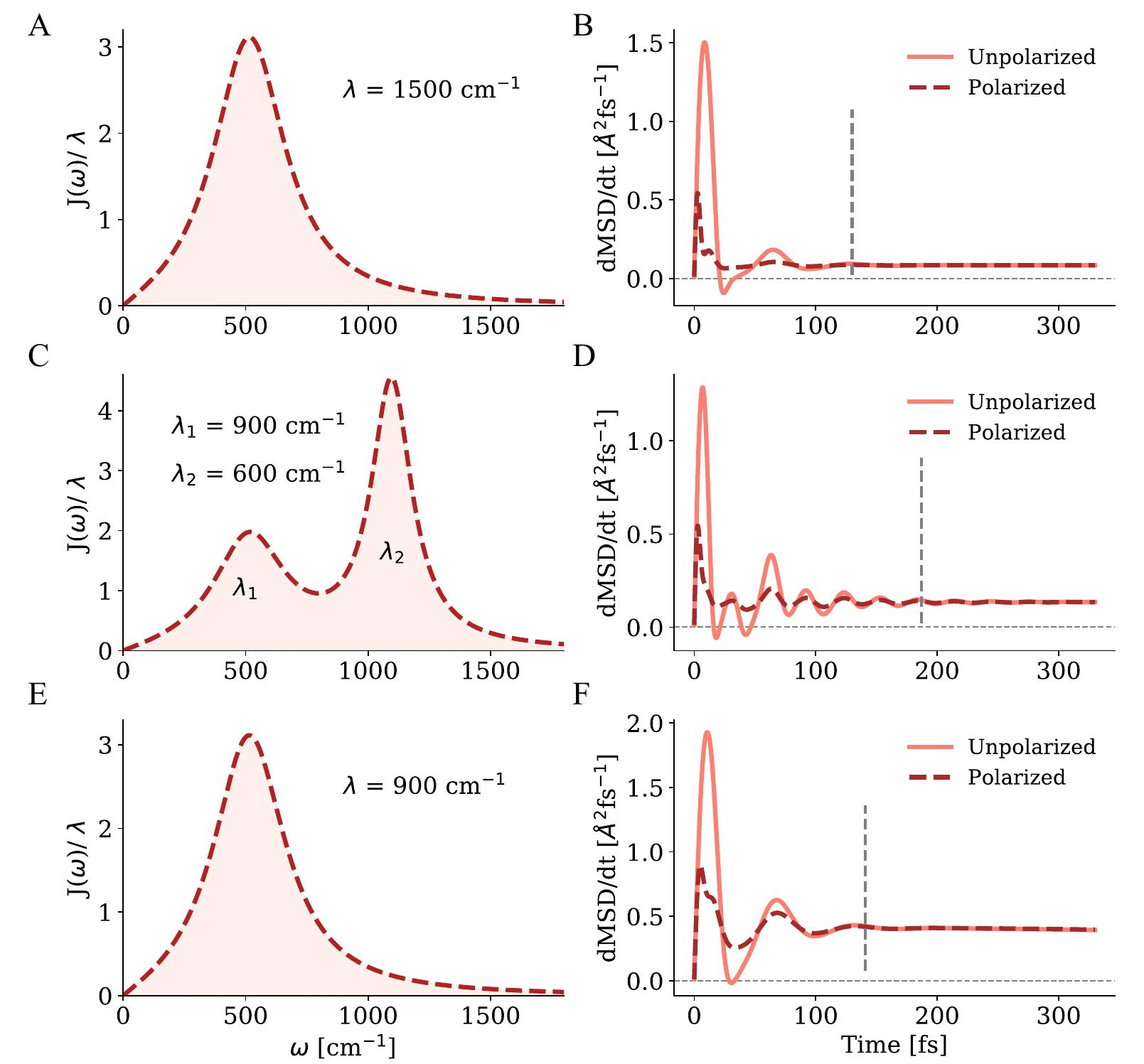}}
\end{center}
\vspace{-24pt}
\caption{\label{fig:cut} Comparison of polaron formation times for three types of spectral density. The spectral density in panel (A) consists of only a single low-frequency peak with the full reorganization energy, $\lambda = 1500\ \rm cm^{-1}$. Panel (B) shows that this spectral density leads to a polaron formation time of $\sim 130$ fs. The spectral density in panel (C) consists of a low- and high-frequency contribution with the reorganization energy distributed in a 3:2 ratio, respectively. Panel (D) reveals that this spectral density leads to a polaron formation time of $\sim 180$ fs. In panel (E), we adopt the same spectral density as in panel (A), albeit with a reduction in its reorganization energy consistent with that of the low-frequency peak in the spectral density of panel (b), i.e., $\lambda = 900\ \rm cm^{-1}$. Panel (F) shows that this spectral density leads to a polaron formation time of $\sim 140$ fs. These results show how varying the portion of reorganization energy across the high- and low-frequency regions of the spectral density tunes polaron formation timescale.}
\vspace{-4pt}
\end{figure}

Given that the width of the spectral density dictates polaron formation timescales, one might anticipate that the total polaron formation timescale in the doubly peaked case might be dominated by the faster timescale set by the broader spectral distribution. However, we find that it is instead close to a weighted average of the two timescales, with the weights being determined by the reorganization energy of each spectral density peak. To illustrate this point, we calculate the polaron formation timescales for each of the three cases with doubly peaked Lorentzian distributions while varying the distribution of reorganization energy between low- and high-frequency regions (see Fig.~\ref{fig:comp1}). We find that shifting the reorganization energy to the narrow-width region (cases 1 and 2) increases the polaron formation timescale. If both low- and high-frequency regions have similar widths, the polaron formation timescale is independent of the reorganization energy distribution among regions (see case 3). We also consider a fourth case analogous to case 1 with a single difference: we separate the high- and low-frequency regions of the spectral density even further. This case allows us to test the effect of the central frequency of each of the two Lorentzian contributions to the spectral density We find the polaron formation timescale does not change significantly in this case (see Fig.~\ref{fig:comp2}.) This result reinforces the idea that the distribution of reorganization energy between the narrow and wide regions of the spectral density plays a decisive role in determining polaron formation timescales.

To further support our argument that the narrow high-frequency region in the spectral density is primarily responsible for slowing down polaron formation, we show a comparison of three specific spectral density configurations and their corresponding polaron formation timescales. First, we consider the `low' case from Table~\ref{table2}, where the total reorganization energy is concentrated in the low-frequency region (Fig.~\ref{fig:cut}-A). This results in a polaron formation timescale of $\sim130$~fs. In the second case, we redistribute the reorganization energy between the low- and high-frequency regions in a 3:2 ratio (Fig.~\ref{fig:cut}-C). In this case, the timescale increases to $\sim185$~fs. Finally, we remove the high-frequency contribution entirely and reduce the total reorganization energy accordingly (Fig.~\ref{fig:cut}-E), leading to a polaron formation timescale of $\sim140$~fs.

These results clearly demonstrate that the presence of a narrow-breadth, high-frequency component, primarily associated with optical phonon modes, significantly slows down polaron formation. Therefore, by engineering or suppressing these narrow high-frequency modes, it may be possible to reduce the polaron formation timescale. In summary, this section emphasizes that both the width of the spectral density (i.e., phonon density) and the distribution of reorganization energy among these phonon modes are key factors in determining the polaron formation timescale. Adjusting either parameter offers a pathway to control and potentially accelerate or decelerate polaron formation.

\bibliography{ref_main.bib}